\def\year{2015}
\documentclass[letterpaper]{article}
\usepackage{aaai}
\usepackage{graphicx}
\usepackage{epstopdf}
\usepackage{subfigure}
\usepackage{times}
\usepackage{helvet}
\usepackage{courier}
\usepackage{amssymb,amsmath}
\usepackage{color}
\usepackage{url}

\newtheorem{observation}{Observation}
\begin{document}
\title{Attention Inequality in Social Media}
\author{
Linhong Zhu and Kristina Lerman\\
USC Information Sciences Institute\\
4676 Admiralty Way\\
Marina Del Rey, CA 90292\\
\{linhong,lerman\}@isi.edu}
\maketitle
\begin{abstract}
Social media can be viewed as a social system where the currency is attention. People post content and interact with others to attract attention and gain new followers. In this paper, we examine the distribution of attention across a large sample of users of a popular social media site Twitter. Through empirical analysis of these data we conclude that attention is very unequally distributed: the top 20\% of Twitter users own more than 96\% of all followers, 93\% of the retweets, and 93\% of the mentions. We investigate the mechanisms that lead to attention inequality and find that it results from the ``rich-get-richer" and ``poor-get-poorer" dynamics of attention diffusion. Namely, users who are ``rich'' in attention, because they are often mentioned and retweeted, are more likely to gain new followers, while those who are ``poor'' in attention are likely to lose followers. We develop a phenomenological model that quantifies attention diffusion and network dynamics, and solve it to study how attention inequality grows over time in a dynamic environment of social media.
\end{abstract}
\section{Introduction}
Inequality is a pervasive social phenomenon. For example,  income and wealth are not evenly distributed in modern society, but instead, concentrated among few individuals. The richest 20\% in the United States own more than 80\% of the country's total wealth, while the poorest 20\% own just a fraction of one percent~\cite{Ariely11inequality}.
Inequality does not just violate our sense of fairness~\cite{Ariely11inequality}, but it also reduces economic opportunities, and has been linked to negative social outcomes, such as poor health~\cite{Kawachi99health} and high crime rates~\cite{Kelly00crime} in societies with high levels of inequality. While economic inequality has been widely discussed recently, inequalities are rampant in other domains, including science and entertainment, where a few individuals receive disproportionate share of attention and the benefits it brings.
In science, specifically, attention can be measured by the number of citations that papers (and scientists who publish them) receive. The distribution of the number of citations is extremely unequal, with some papers receiving thousands of citations, while many others are seldom cited~\cite{Allison80,Klamer02}. Researchers have long speculated about the origins of inequalities. Current consensus holds that inequality cannot be explained by variation in quality of the work or individual talent~\cite{Adler,Salganik06,Lariviere09}. Rather, it is the result of the underlying social processes, such as the ``cumulative advantage'' or the ``rich-get-richer'' effect, which brings greater recognition to those who are already distinguished~\cite{Merton68,Allison82}.

In online social media attention is also non-uniformly distributed. Some user-generated photos and videos are viewed, liked, or shared orders of magnitude more times than the rest. Psy's Gangnam Style video, for example, was viewed over 2 billion times on YouTube, while the average video has fewer than 100 thousand views. Attention paid to people is also highly unequal. A few Twitter users have tens of millions of followers, while the majority have just a handful of followers. These popular users enjoy an unparalleled advantage: things they say on Twitter, from product endorsements to personal opinions, are seen by millions, making them potentially far more influential than ordinary users.

In this paper, we carry out quantitative analysis of data from the microblogging service Twitter to characterize and quantify attention inequality in social media.
We identify a set of almost 6,000 users and monitor their activity over a period of several months.
We show that attention is very unequally distributed: the top 20\% of Twitter users in our sample own 97\% of all followers, 93\% of retweets, and 93\% of mentions.
By observing how the monitored users allocate their attention over time, via following, mentioning or retweeting other users, we are able to study the processes leading to attention inequality. We empirically demonstrate that the attention Twitter users receive by being mentioned or having their posts retweeted by others results in them gaining new followers. In contrast, users with fewer followers are retweeted less frequently and are more likely to lose followers. We construct a phenomenological model of attention diffusion where these two processes --- mentioning and retweeting --- result in changes in the structure of the underlying follower network.  After parameterizing the model, we solve it to study how attention and inequality evolve over time. Our model produces a ``rich-get-richer''  and ``poor-get-poorer" dynamic, whereby users with more followers, i.e., Twitter celebrities, are mentioned and retweeted more frequently, which results in them gaining even more followers. Our model is able to identify users who will gain more attention, to predict the evolution of the follower network and attention inequality.
%
%
%

There are consequences to high attention inequality, few of which have been explored in the context of social media.
High inequality focuses the attention of the online community on relatively few users, allowing them to disproportionately benefit from the attention they receive. Companies compensate social media celebrities for endorsing their products and messages. Ordinary users receive no such pay-offs. Attention is the currency of social media and, similar to other currencies, it is finite. When everyone is retweeting a popular celebrity or watching a viral video, it means that other people are not being retweeted, and other videos are not being watched. Thus, inequality diminishes the diversity of content and view points to which people are exposed.
But inequality may also have hidden benefits. When every single person had watched the same video, or read the same story, it gives people a common topic for conversation (and tweeting) and a common vocabulary with which to discuss. Thus, inequality reinforces social identity.

Our work raises a number of questions, including why do users tolerate extreme attention inequality? How does it affect user behavior? Does it reduce their engagement with Twitter, or does it inspire them to be more active in gaining attention? And not least, should Twitter, and other social media sites, do something to ensure attention is more equitably distributed? We hope that our work stimulates research community to address these questions.

%

%
%

%
%
\section{Quantifying Attention Inequality}
\label{sec:attention}
In this section we quantitatively characterize the attention received by users and content they post on Twitter. Although we focus on Twitter, our definitions of attention generalize to other social media platforms.

\subsection{Data Description}
We collected data from Twitter over a period from March 2014 to Oct 2014 using the following strategies. First, we randomly selected 5600 seed users whose user ids range from [0,760000000].
We then used the Twitter API to periodically query the friends of seed users (i.e., whom the seed users were following), which resulted in a dynamic network where each edge had a time stamp with its creation and/or deletion date. Next, we collected profile information for all the seed users and their friends, as well as timeline tweets.  The number of tweets gives that user's activity, or engagement, level. We tracked the retweet/reply/mention status of each tweet from its raw Jason object returned by Twitter. The number of retweets gives the number of times users have shared a particular tweet with their own followers. As we argue later, this is an important measure of attention. For the seed users, we further monitored their temporal profiles by querying user profile weekly during the data collection period.

\begin{figure}
  \centering
 \subfigure[]{\label{subfig:distfollower}\includegraphics[width=0.492\columnwidth]{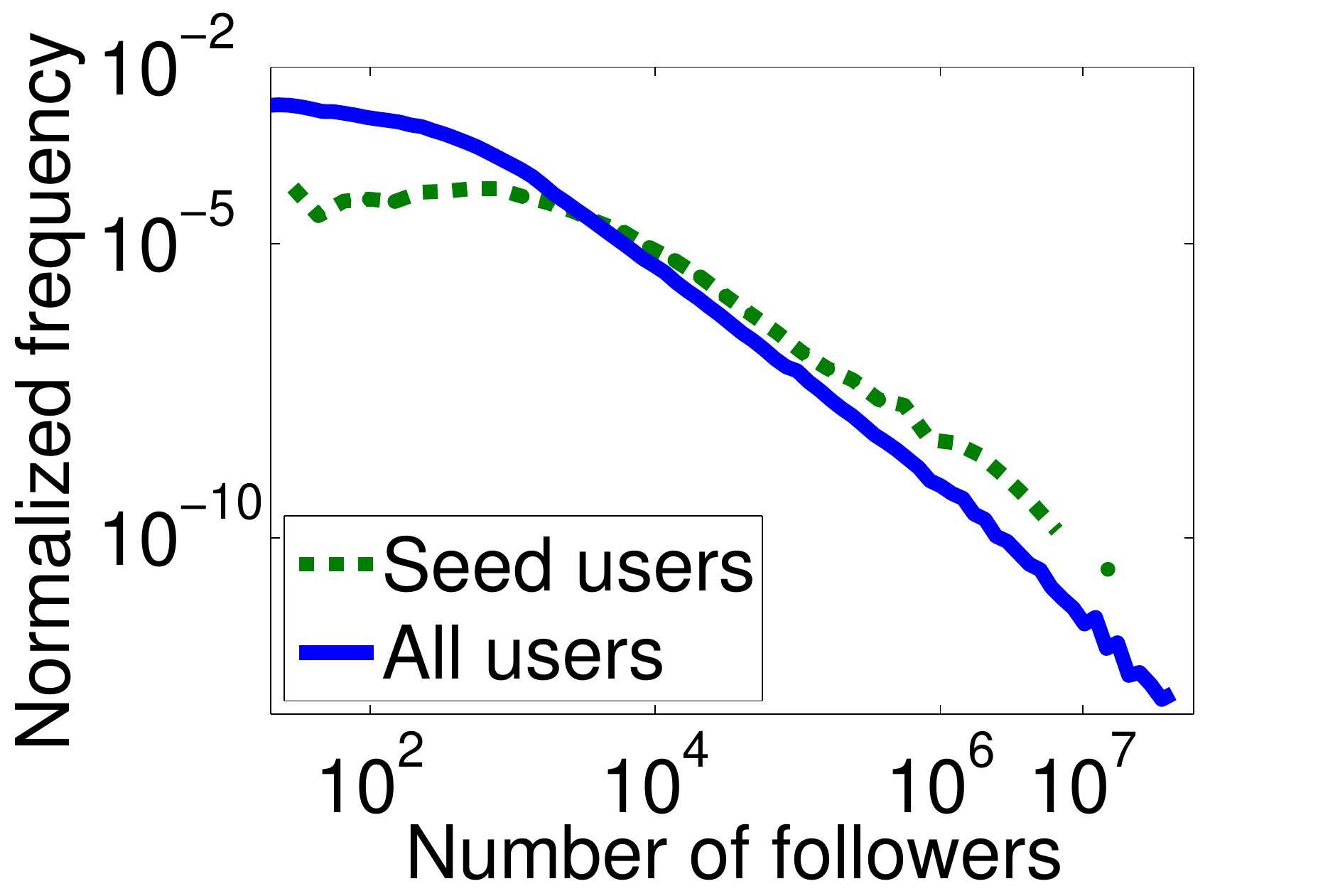}}
 \subfigure[]{\label{subfig:distfriend}\includegraphics[width=0.492\columnwidth]{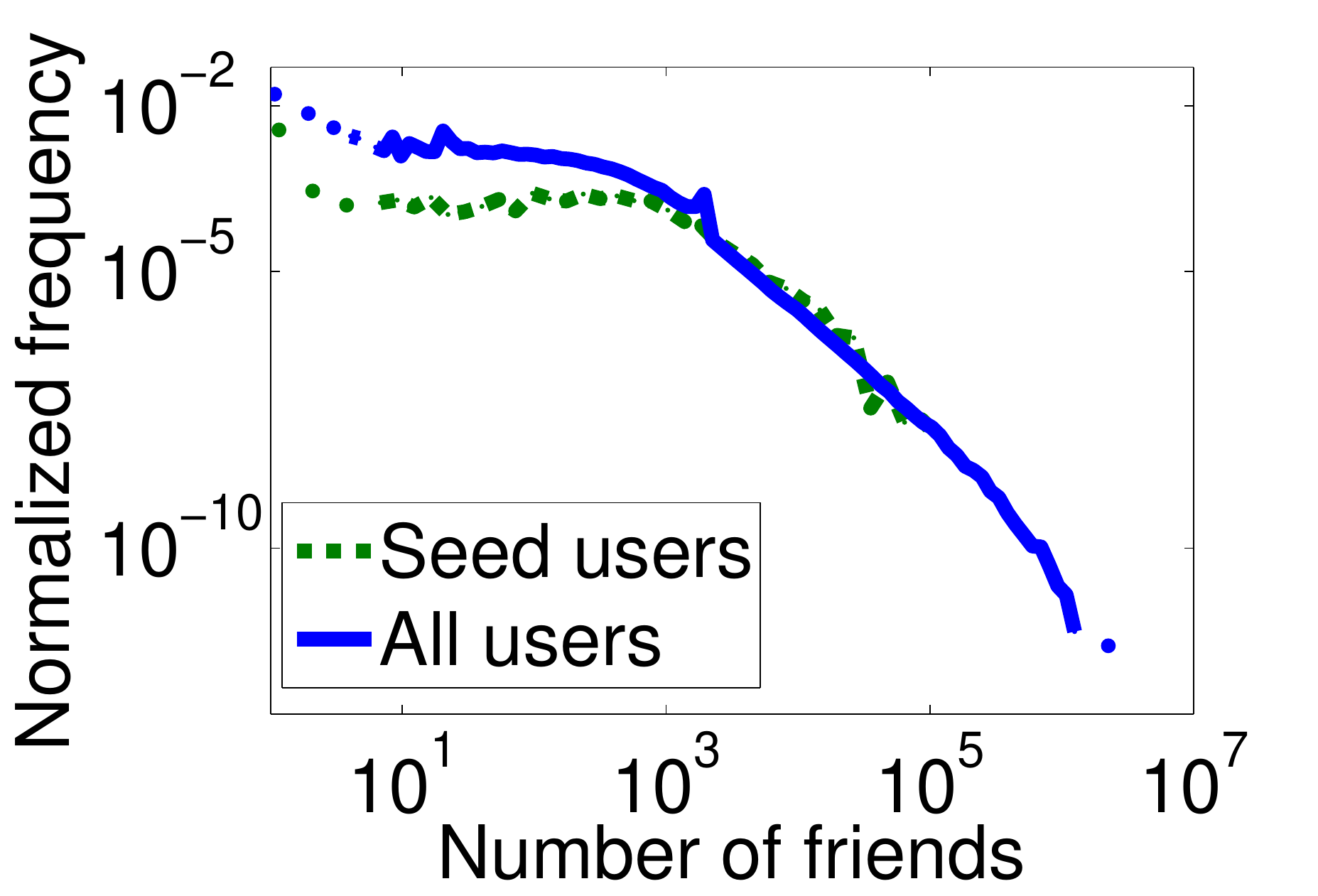}}
  \subfigure[]{\label{subfig:disttweet}\includegraphics[width=0.492\columnwidth]{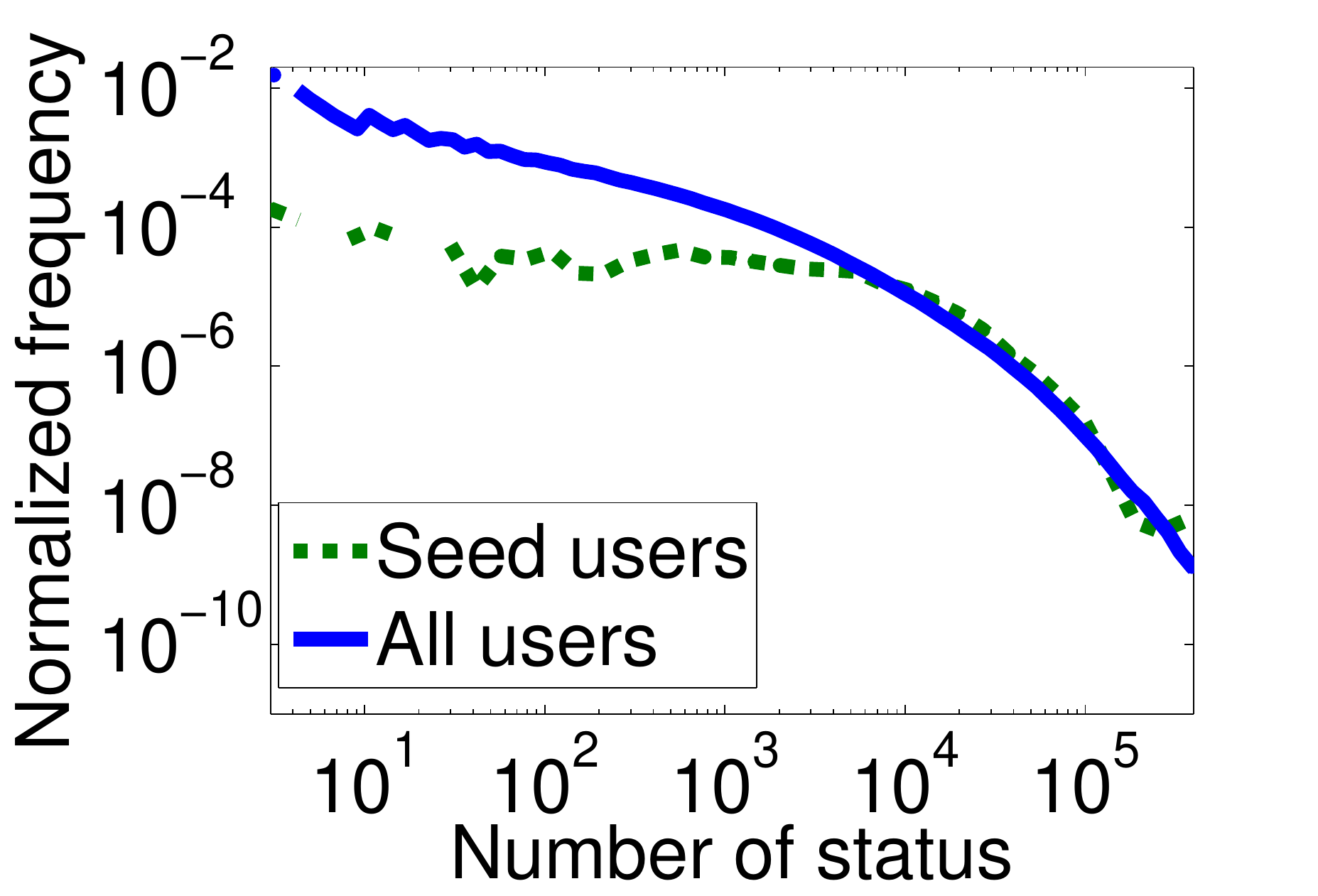}}
 \subfigure[]{\label{subfig:distfavor}\includegraphics[width=0.492\columnwidth]{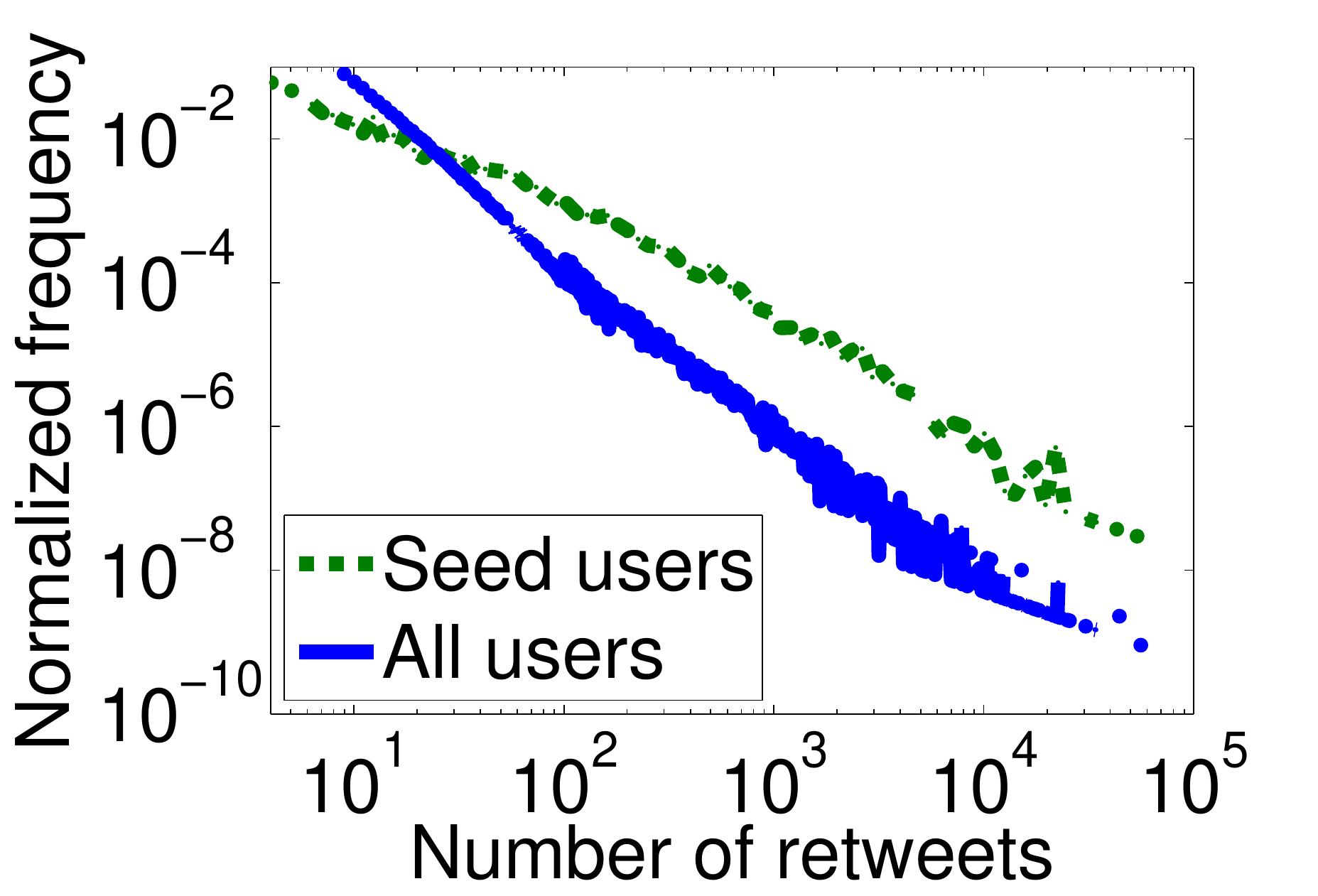}}
  \caption{Distribution of the numbers of (a) followers, (b) friends, (c) status updates posted, and (d) retweets received for seed users and all users. We logarithmically bin data for smoothing.}\label{fig:distribution}
\end{figure}

The data collection process resulted in 23,831,568 tweets and a dynamic network with 1,944,383 users and 17,869,415 edges.
To provide more details about statistics of our data, we analyze the distribution of the numbers of followers, friends, status updates (user activity), and the times that the tweets of the seed users, and all users in our data, were retweeted. Note that we use `status updates' to refer to the aggregate tweets posted by user since account creation, and we use `tweets' to refer to the tweets user posted during a specific time period. These distributions, plotted in Figure~\ref{fig:distribution}, have a characteristic long-tailed form, where a few users have extremely large numbers of friends, followers or posted tweets, while many users have few friends, followers or posted tweets.
Such distributions are generally associated with high inequality.

Except for activity (number of posted tweets) and number of retweets, the distributions for seed and all users are very similar, which indicates success of our random seed user selection strategy. The small differences between the distribution of number of tweets among seed users and among all users arise because the profile information of seed users is obtained from raw Jason objects of tweets, and thus users who never post any tweets cannot be selected as seed users in our data collection. Due to these differences, the distribution of number of retweets among seed users, also slightly differs from that among all users.

%

\subsection{Attention Inequality}

%

How much attention do Twitter users receive? Certainly, it makes sense to use the number of followers as a measure of attention: the more followers you have, the more people will see the messages you post. However, there is also a wide distribution in user activity, and a person with a million followers who never posts anything will not receive any attention. Therefore, we propose two additional measures of attention --- the number of times a user's posts are retweeted, or reshared by others, and the number of times a user is mentioned by others. These measures of attention are related: on average, the number of times a user's post is retweeted  or the number of times a user is mentioned is proportional to the number of followers the user has. However, using the total number of retweets allows us to include variation in user activity in a measure of received attention; while using the total number of mentions includes effects of celebrity. Since we have the network, profile and activity information only for seed users, we analyze how these seeds users allocate their attention to other users.

\begin{figure}
  \centering
  \subfigure[]{\label{subfig:lorenzattention}\includegraphics[width=0.492\columnwidth]{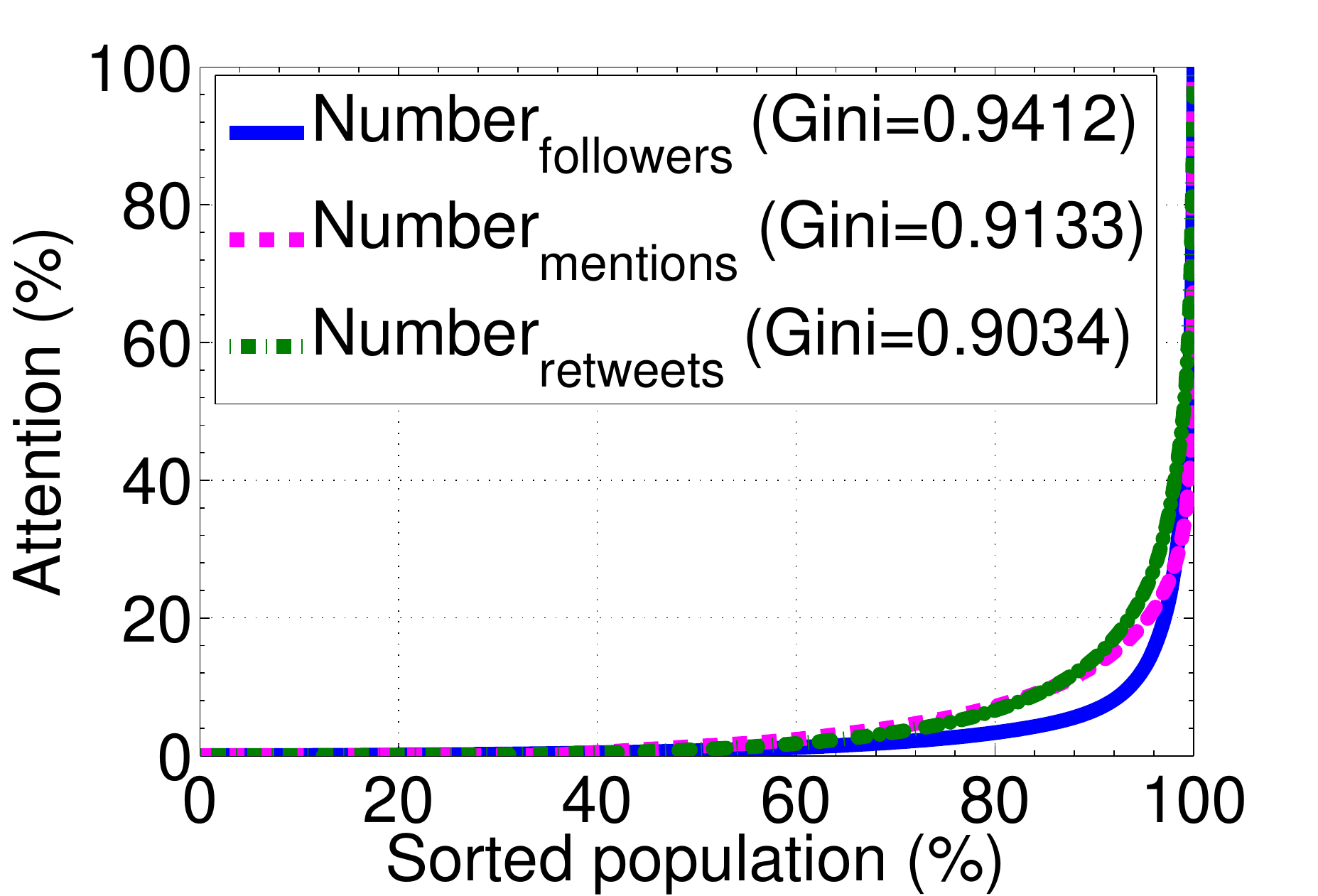}}
  \subfigure[]{\label{subfig:lorenzactivity}\includegraphics[width=0.492\columnwidth]{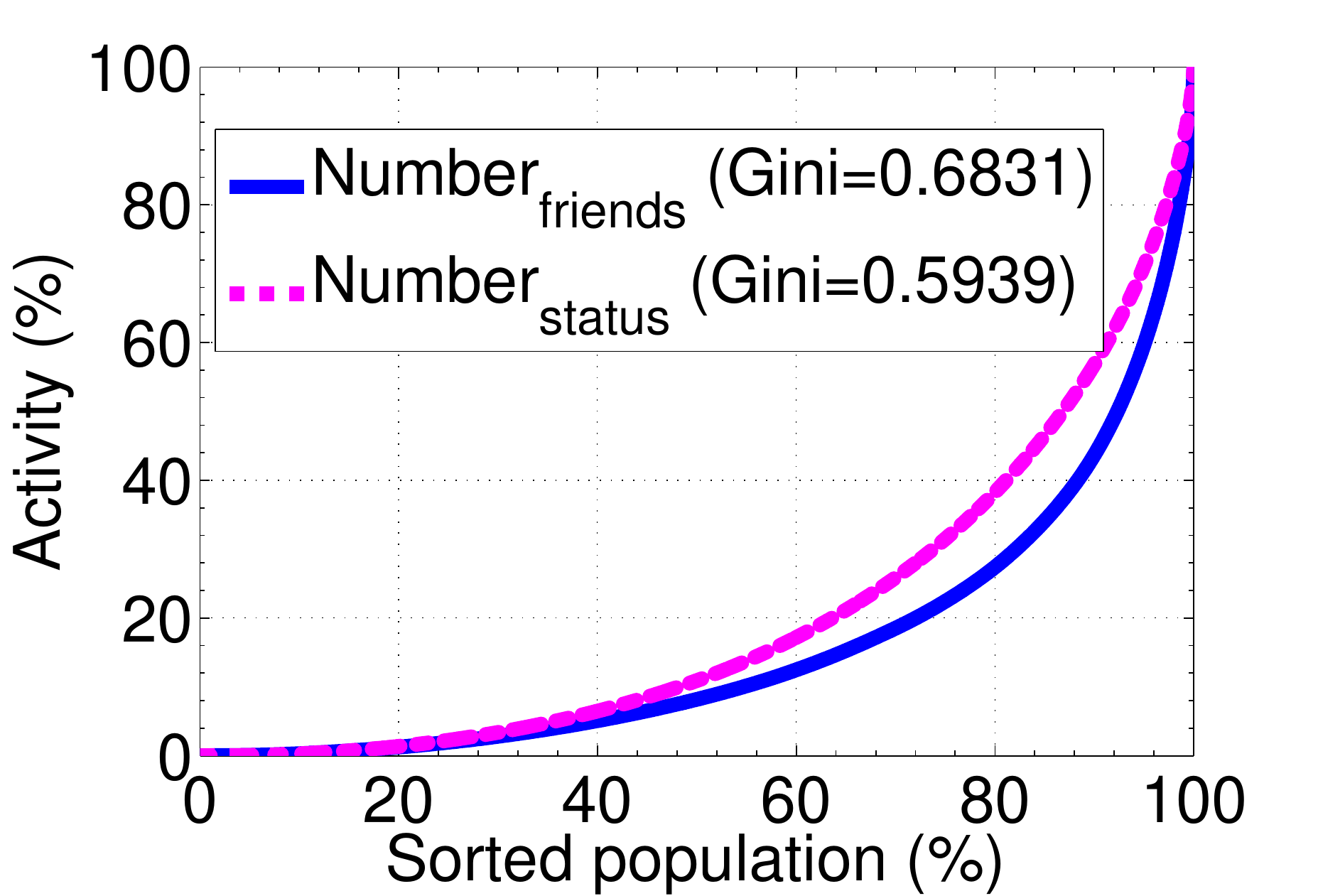}}
  \caption{Lorenz curves and Gini coefficients of inequalities of (a) attentions and (b) activities among seed users.}\label{fig:gini}
\end{figure}

Inequality of a distribution can be characterized using the Lorenz curve, which graphically represents the cumulative distribution function of the empirical probability distribution. In other words, it plots the cumulative share of value, e.g., number of followers, as a function of the cumulative percentage of the population who have at least that many followers. Perfect equality is a straight $y=x$, in which the bottom $x$\% of people have $y$\% of the followers. However, when inequality exists, the Lorenz curve is highly skewed. Figure~\ref{subfig:lorenzattention} shows the Lorenz curves for the number of followers of seed users, the number of times they were mentioned, and the number of times their posts were retweeted. It shows attention unequally distributed: the top 1\% of seed users account for approximately 50\% of the total retweets, 60\% of the total followers, and 66\% of the total mentions.


Another way to characterize the inequality of a distribution is with the Gini coefficient~\cite{gini_concentration_1997}. Given a sample of data $x_1, \cdots x_i$, the Gini coefficient can be calculated from the following equation:
\begin{equation}\label{equ:gini}
g=\frac{\sum_i\sum_j|x_i-x_j|}{2N^2\overline{x}}
\end{equation}
Gini coefficient is zero, meaning there is perfect equality, when all values in a sample are the same, and it attains a value one, meaning maximum inequality, when one number in the set is nonzero and the rest are zeros. Gini coefficient also gives the area between the Lorenz curve of a distribution and the line of perfect equality. Gini coefficient of the number of followers in Fig.~\ref{subfig:lorenzattention} is $g=0.9412$, for the number of mentions it is $g=0.9133$, and for the number of retweets it is $g=0.9034$. The Gini coefficient of the much discussed income inequality is less than 0.5 for the United States, and the much larger wealth inequality has Gini coefficient of around 0.8.\footnote{\url{http://en.wikipedia.org/wiki/List_of_countries_by_distribution_of_wealth}} Attention inequality on Twitter is staggering.  The vast majority of users do not receive any attention, and the top 1\% of users get far more attention than the bottom 99\% combined! What then drives user engagement and activity on Twitter? A measure of user engagement is the number of friends they follow and the number of tweets they post.
Figure~\ref{subfig:lorenzactivity} shows the Lorenz curves of these distributions for the seed users in our data. Results indicate that user engagement is also unequally distributed, with Gini coefficient of $g=0.6831$ for the number of friends and $g=0.5939$ for the number of tweets. Although still unequal, user activity is more equally distributed than attention.



\subsection{Dynamics of Inequality}
\begin{figure}[thb]
  \centering
  \includegraphics[width=\columnwidth]{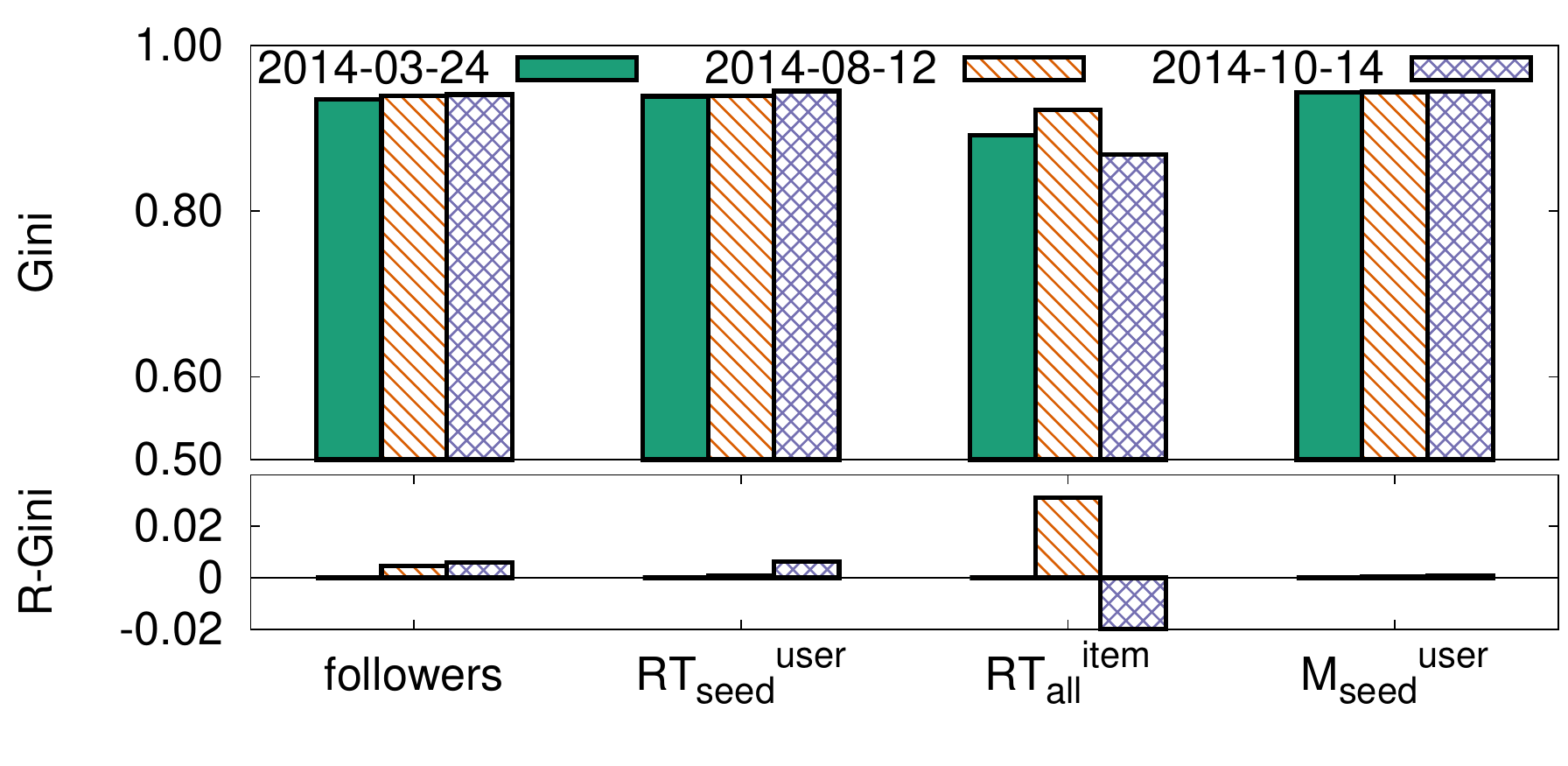}
  \includegraphics[width=\columnwidth]{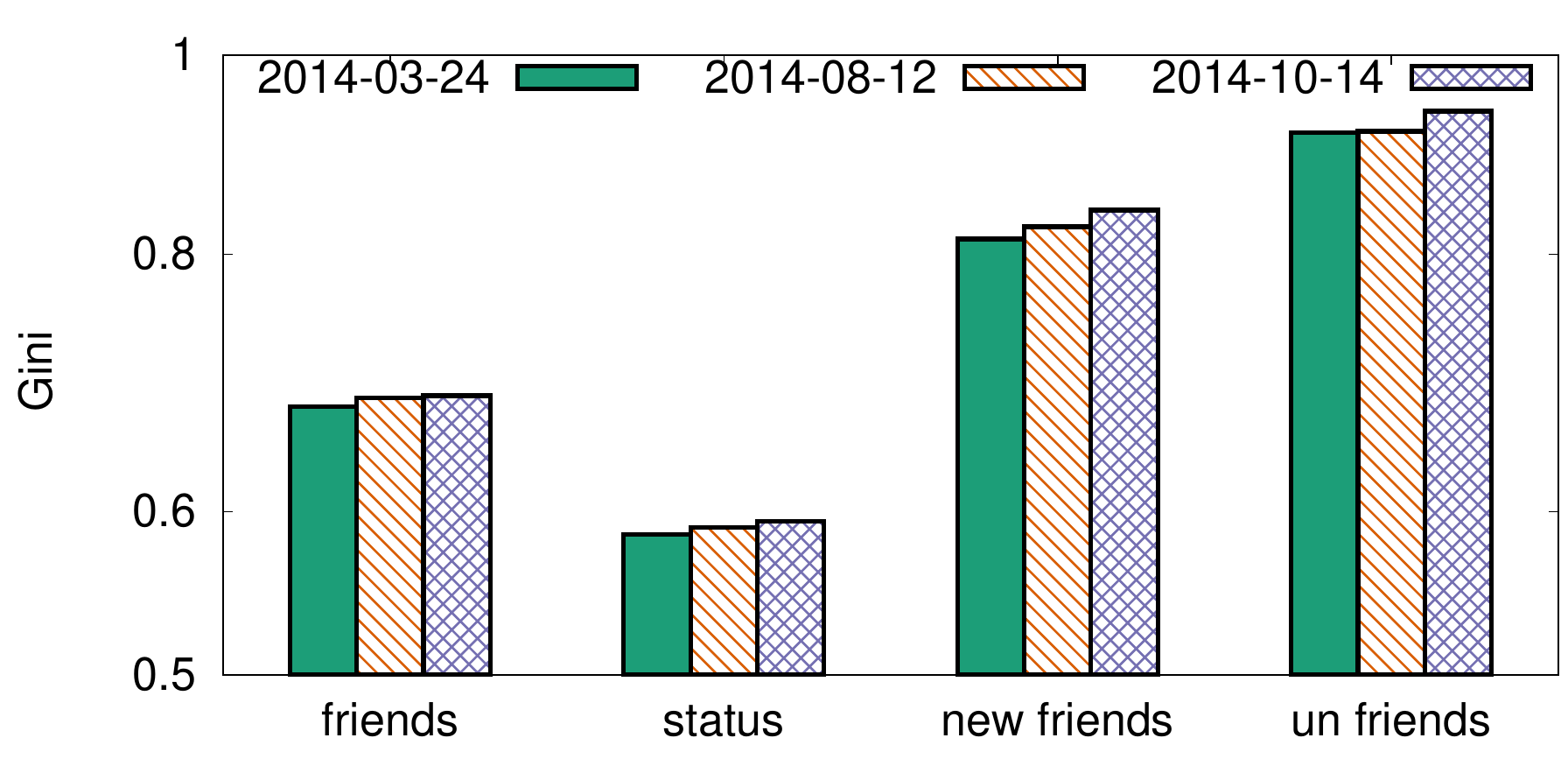}
  \caption{Change in inequality users over time. Here $RT_{\texttt{all}}^{\texttt{item}}$ indicates the number of re-tweets a tweet that is posted by any seed user received from all users in one week period; while $RT_{seed}^{user}$/$M_{seed}^{user}$ denotes the number of retweets/mentions a user received from seed users in one week period at three points in time. }\label{fig:dynamicgini}
\end{figure}

Is high inequality simply an artifact of Twitter's young age and will correct itself over time as users distribute their attention more equitably? Or is the inequality increasing over time?
Figure~\ref{fig:dynamicgini} shows how inequality changed over time during data collection period.
The plot shows the Gini coefficients of different measures of attention at three points in time. The inequality of the number of followers, and the number of retweets of a user ($RT_{seed}^{user}$) are increasing over time; while attention to content, given by the number of times different items are retweeted by all users ($RT_{all}^{item}$) fluctuates over time. The attention inequality in terms of number of mentions, increases over time, but the margin is too small to be significant.
The plot also shows that the attention that items posted on Twitter receive ($RT_{all}^{item}$) is somewhat more equitably distributed than the attention paid to users.

In addition, we calculate inequality of user activity at different points in time, which includes the number of friends, number of status updates, number of unfriends (i.e., number of lost followers), and number of new friends gained at different time periods. The results are plotted in the lower part of Figure~\ref{fig:dynamicgini}. Note that we use number of status updates to denote the aggregate number of tweets users made since account creation. These results indicate that activity inequality also increases over time, which indicates that both new content and new links are created by a small set of super-active users. Also, activity inequality, while high, is substantially less than attention inequality.

These findings suggest that there are processes that drive attention to be concentrated on few individuals at the exclusion of others, even though user engagement is far more equally distributed. In the next section we examine these questions to construct a model of the dynamics of attention.


\section{Attention Diffusion Model}
To help explain dynamics of attention inequality in online social media, we develop a model of network dynamics via diffusion of attention. This model captures the evolution of the Twitter follower network as users are brought to the attention of potential new followers. As these subsequently link to the original users, the Twitter network grows, bringing even more attention to the original users.

One of the best-known models of network evolution is the BA model~\cite{Barabasi99emergenceScaling} with preferential attachment mechanism. In this model, a new node links to an existing node with probability proportional to its degree. This makes it more likely for high degree nodes to acquire new links, resulting in a ``rich get richer'' phenomenon. 
This is a plausible model for the growth of the Twitter follower network: users who already have many followers are more likely to get attention and attract even more followers.
However, Twitter platform is highly complex and dynamic:  users post messages, which may then be retweeted by others. These retweets create opportunities for new follow links to be formed~\cite{Weng13kdd}. Users are also mentioned by others, which also creates opportunities for new follow links~\cite{Zhu14socialcom}. However, links are often broken as users unfollow others~\cite{MyersWWW2014}. Below we empirically characterize how these factors contribute to the growth --- and the loss --- of attention on Twitter.

\begin{table}
\centering
  \caption{Statistics of classes when users are divided into quintiles based on the number of followers they have. The numbers represent the average number of followers and the average activity of users in each class.}\label{tab:stat}
   \begin{tabular}{|c|c|c|c|}
    \hline
     \multicolumn{2}{ |c|}{ \emph{quin-tile}}& \# \emph{statuses} & \# \emph{followers} \\
     \hline
    $Q_1$ & $1^{st}$ & 14,086 & 29,021\\
    \hline
    $Q_2$ & $2^{nd}$ &7,184 &1,058\\
    \hline
    $Q_3$ & $3^{rd}$ &4,053  &445\\
    \hline
  \end{tabular}
\end{table}

Twitter users are heterogeneous, and popular users who are ``wealthy'' in attention could behave in a qualitatively different manner from the less popular users. To partially control for heterogeneity, we split users into classes according to the number of followers they have. For simplicity, we use five classes, corresponding to quintiles in the number of followers. Table~\ref{tab:stat} reports statistics of the top three quintiles. Users in the first quintile, representing the 20\% of the users with most followers, have over 29K followers on average and tweeted over 14K status updates on average. Users in the second quintile, representing the next 20\% of most popular users, have almost 1K followers on average and posted almost 7K tweets, while users in the third quintile have a little more than 400 followers and posted almost 4K times. Users in the bottom two quintiles did not have much activity and were excluded from analysis. The main idea behind this division is that users within the same quintile will be more homogeneous and similar to each other.

\begin{figure}[thp]
\centering
\subfigure[]{\label{subfig:T2F}\includegraphics[width=0.492\columnwidth]{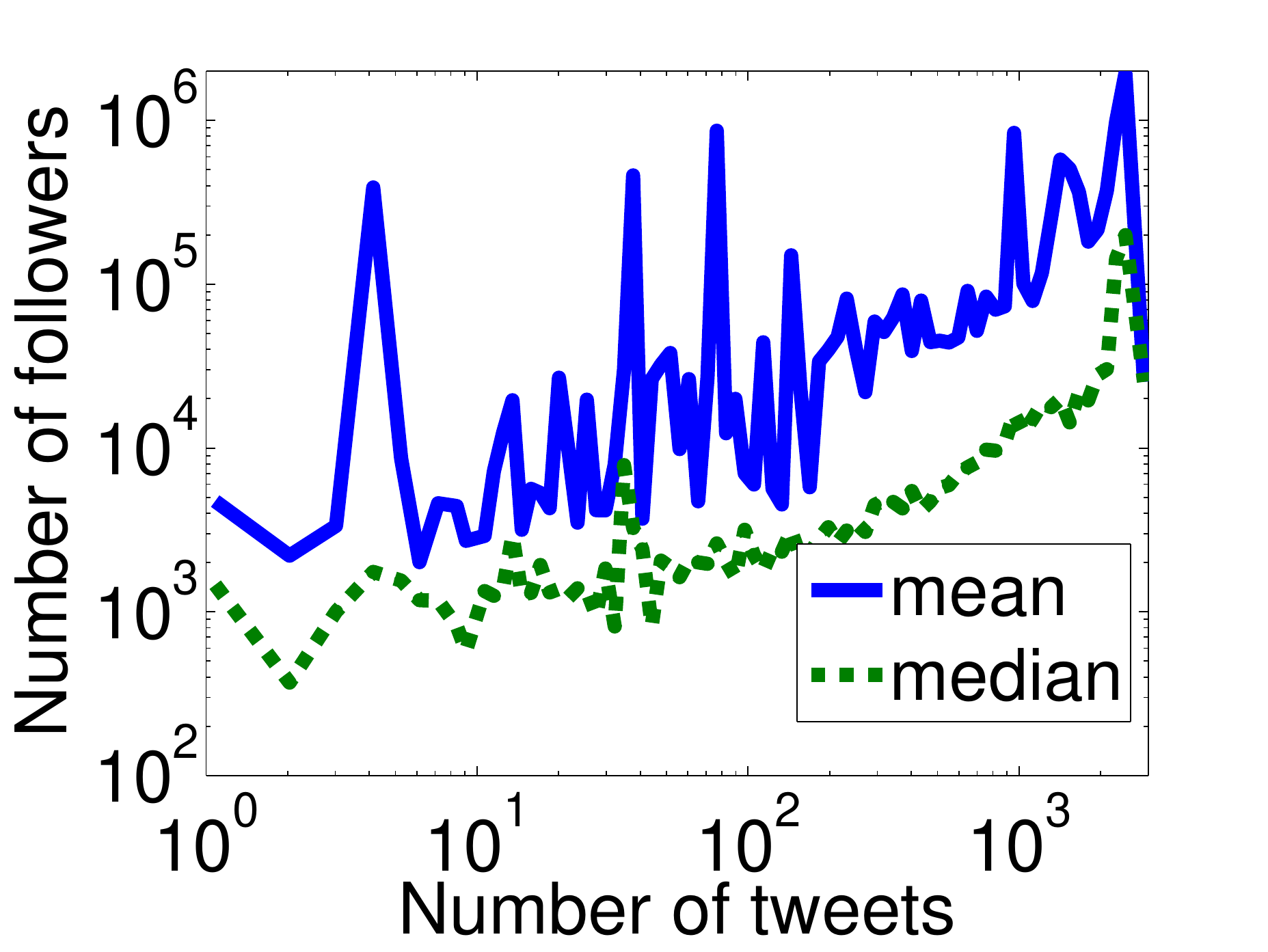}}
\subfigure[]{\label{subfig:T2RT}\includegraphics[width=0.492\columnwidth]{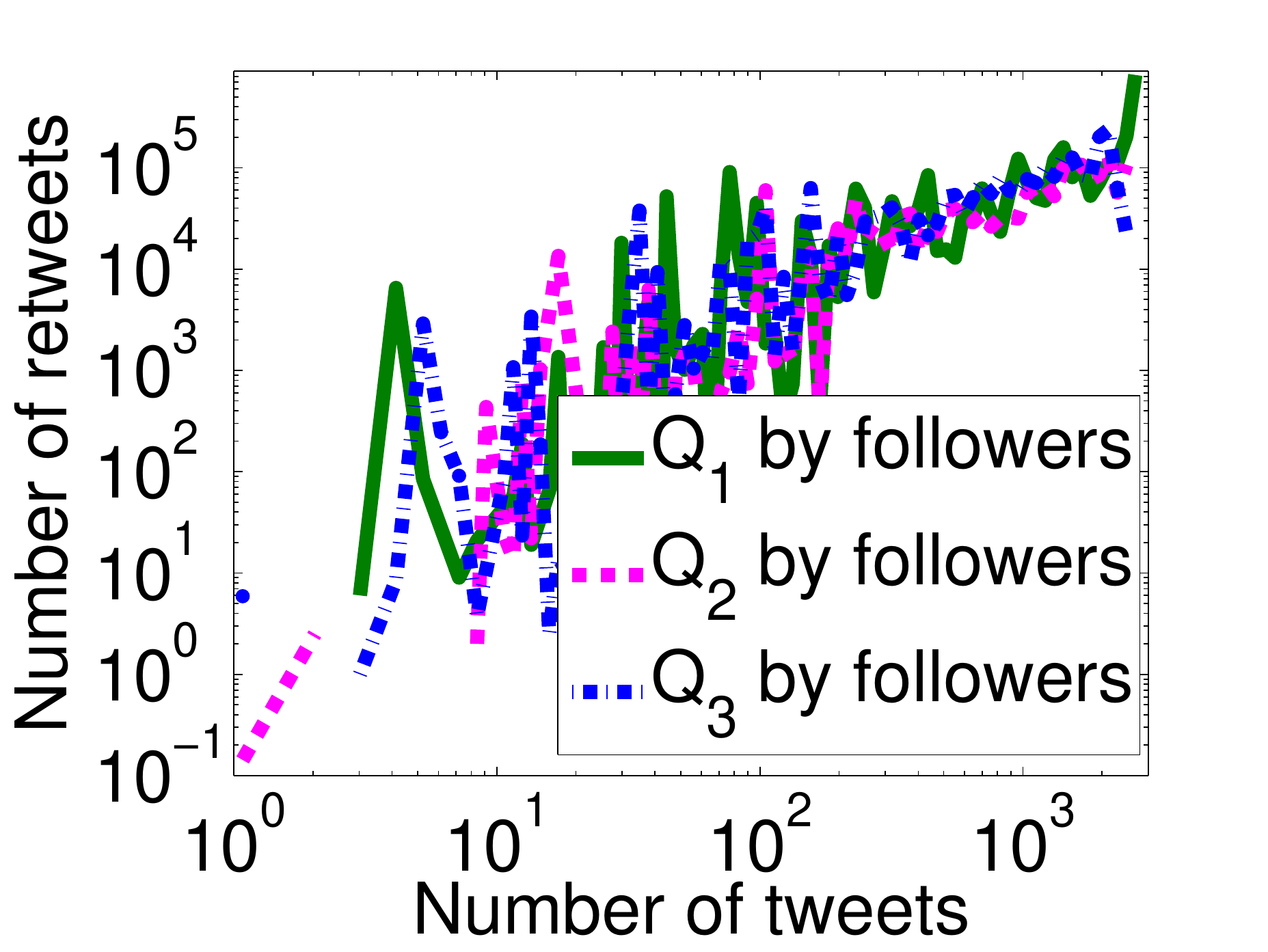}}
\caption{Relation between posting activity and attention ((a) number of followers, (b) number of retweets). More tweets a user posted, more followers and retweets a user received.}\label{fig:post2attention}
\end{figure}

After creating more homogeneous populations, we examine how the attention users receive leads them to gain, or lose, followers. Again, we quantify attention by the number of times users are mentioned and the number of times they are retweeted by their followers. Both of these depend on the number of followers users currently have. While this is not directly under user's control, they can try to increase the number of followers by following other users, and hoping that some of these will result in reciprocal links. Users can also attempt to increase the number of times they are retweeted by posting more tweets, although bursts of tweeting activity may cause followers to classify users as spammers and unfollow them. Below we examine how these factors are correlated with the attention users receive and their probability to gain or lose followers.

\begin{observation} \textbf{\emph{(Effect of tweeting)}}
Users who post more tweets receive more attention. 
\end{observation}

We examine the relationship between posting activity, measured by the number of status updates users tweet, and the number of followers they have. Figure~\ref{subfig:T2F} confirm an earlier finding~\cite{Hodas13icwsm} that activity and number of followers are positively correlated. Posting activity also affects how much attention user's posts receive, which we measure by the number of times the posts are retweeted. Figure~\ref{subfig:T2RT} shows a clear positive correlation between activity and number of retweets for the first three user quintiles. Since each retweet increases user's visibility by bringing his name into the feeds of the retweeting user's followers, it also increases the likelihood of gaining new followers~\cite{Weng13kdd,Zhu14socialcom}. We do not show the plot that evaluates the relationship between posting activity and mentions since that correlation is relatively low 0.0259.

\begin{figure}[!t]
\centering
\subfigure[]{\label{subfig:Fr2F}\includegraphics[width=0.492\columnwidth]{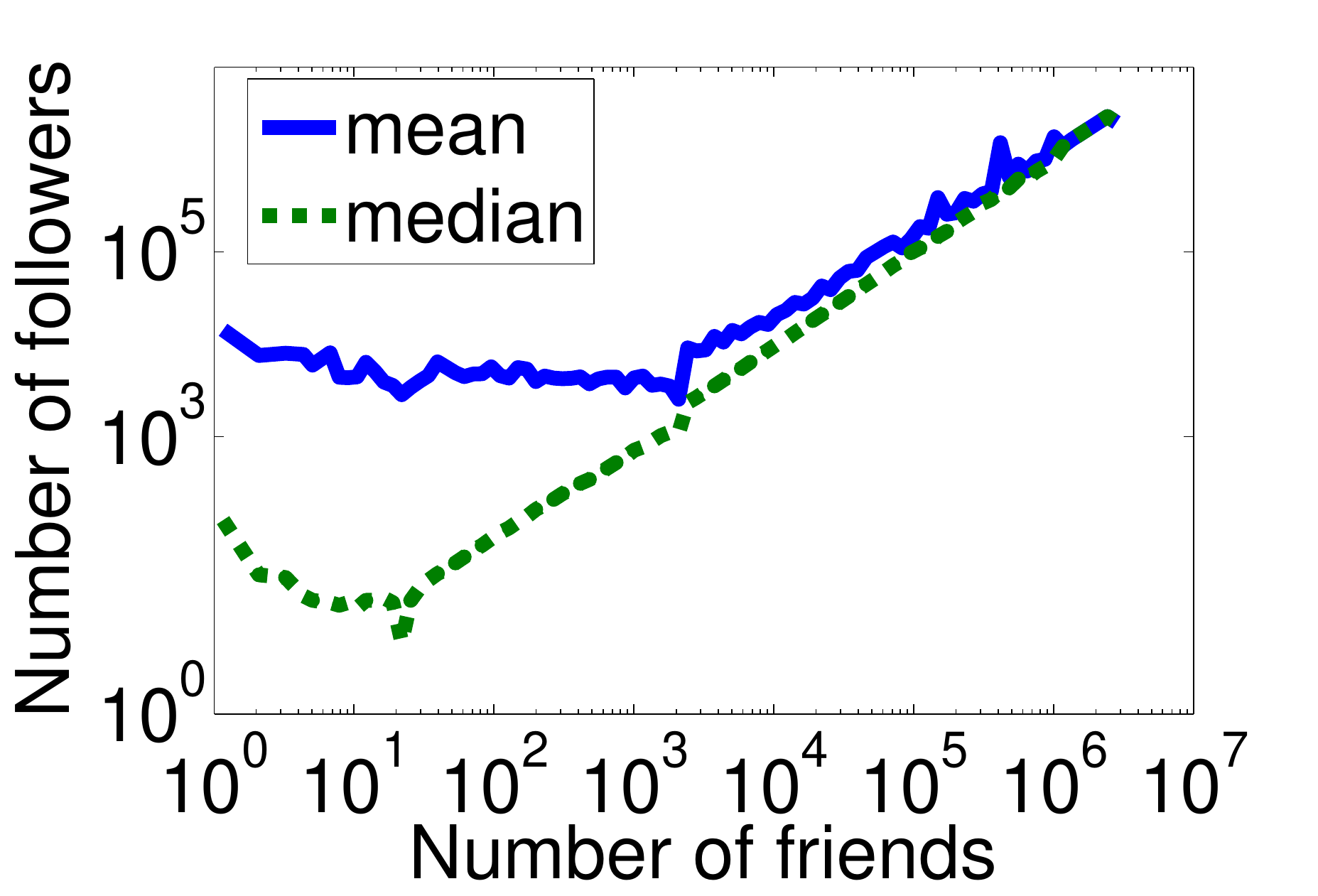}}
\subfigure[]{\label{subfig:Fr2RT}\includegraphics[width=0.492\columnwidth]{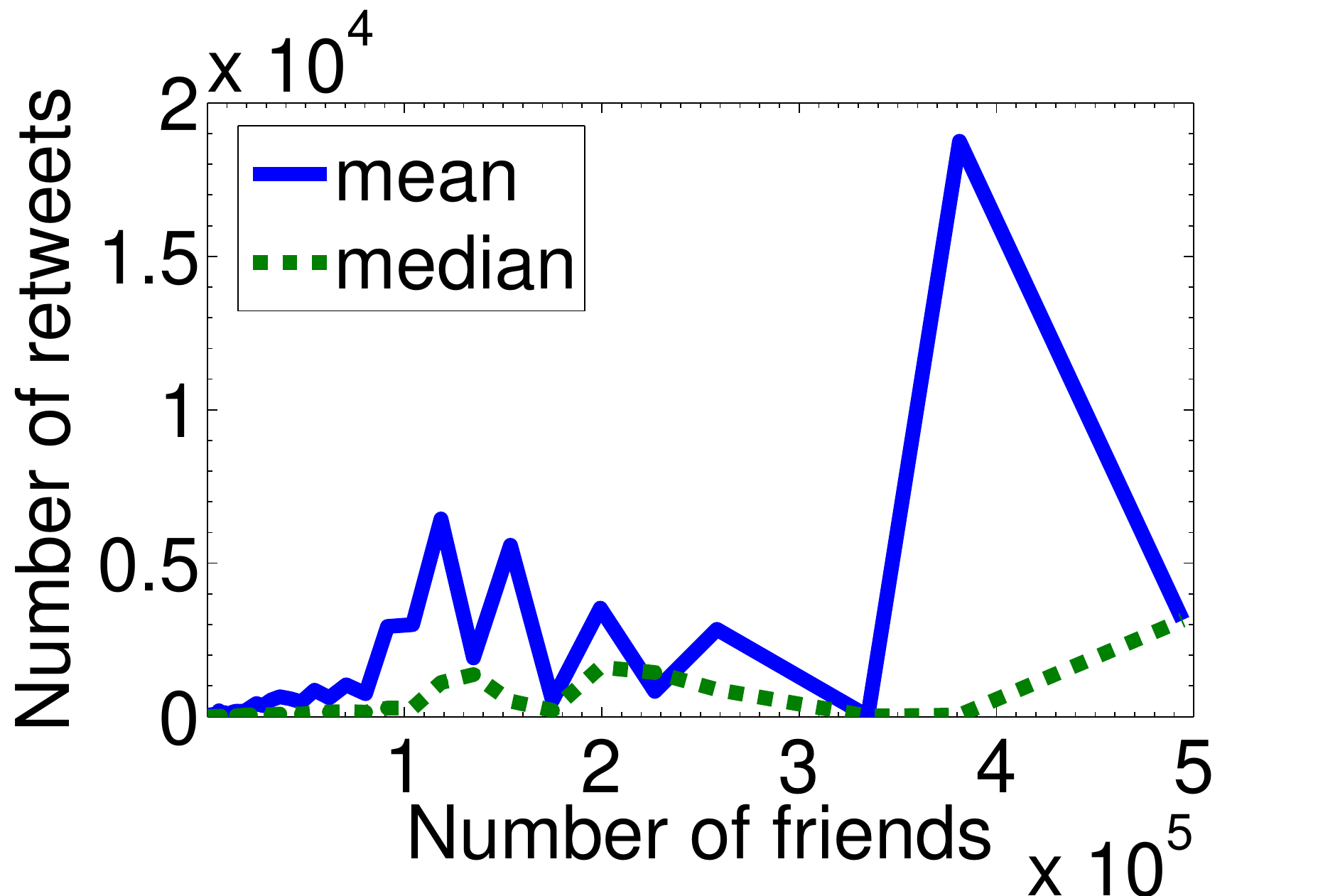}}
\caption{Relation between following activity and attention ((a) number of followers, (b) number of retweets). }\label{fig:factor2re-tweet}
\end{figure}

\begin{observation} \textbf{\emph{(Effect of following)}}
The following others does not necessarily increase attention.
\end{observation}
Next, we examine whether by following more people, users can increase the attention they receive. In general, there is a positive correlation between the number of friends users follow and the number of followers they have (Figure~\ref{subfig:Fr2F}). The exception to this trend are the few outliers who have fewer friends than followers. Most of these users represent organizational accounts, such as YouTube with only 851 friends but 47.6 million followers, or celebrities, such as Lady Gaga with 133 thousand friends but 43.6 million followers. However, as shown in Figure~\ref{subfig:Fr2RT}, there is no strong correlation between the number of friends and the number of retweets user receive. Similarly, the correlation between following activity and number of mentions is only 0.0124. Therefore, following more people does not translate into increased attention.

\begin{observation}\textbf{\emph{(Effect of attention)}}
The more attention users receive, via retweets and mentions, the more followers users will gain.
\end{observation}

\begin{figure}[!t]
\subfigure[]{\label{subfig:R2prob2}\includegraphics[width=0.492\columnwidth]{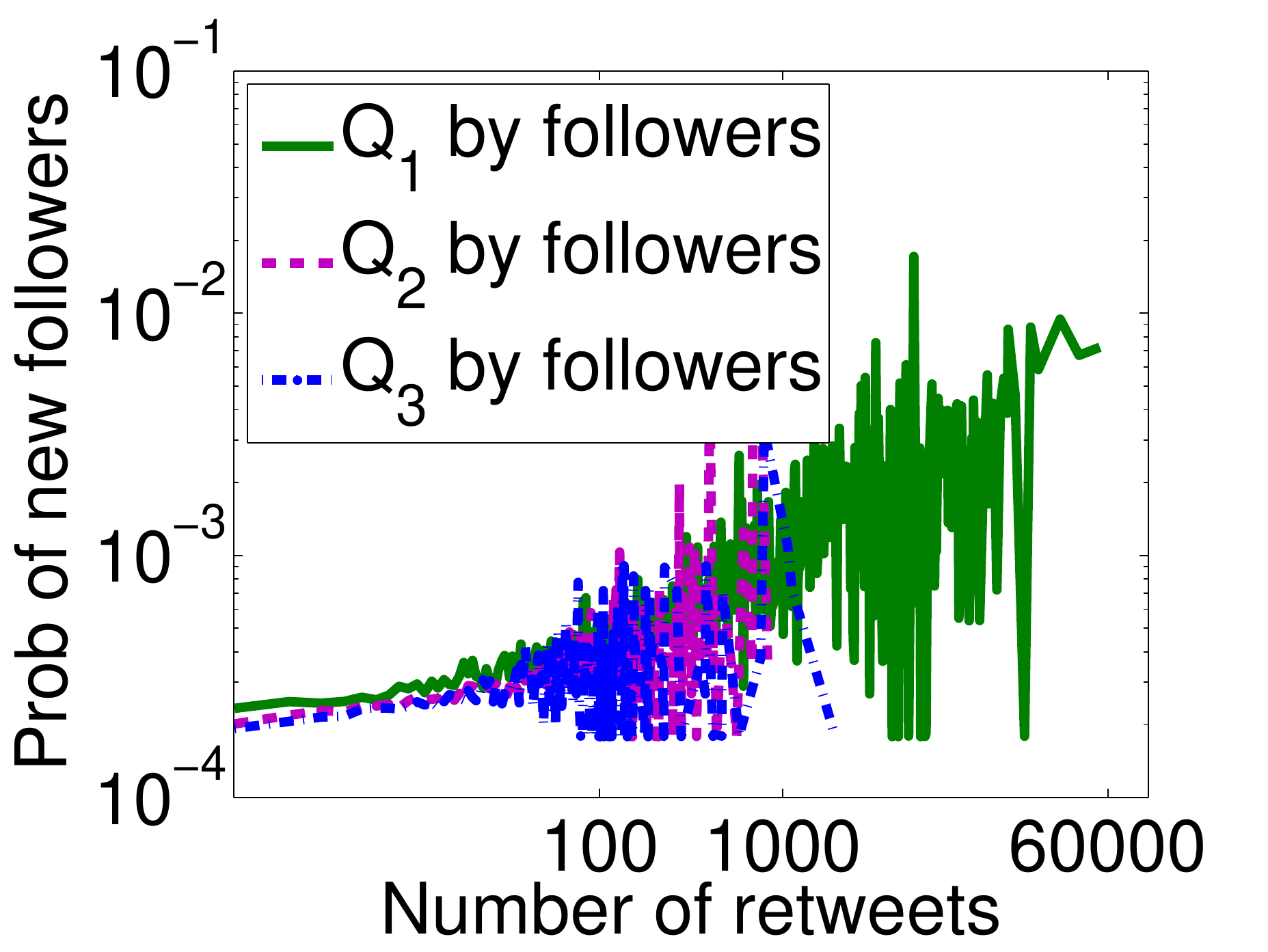}}
\subfigure[]{\label{subfig:M2prob}\includegraphics[width=0.492\columnwidth]{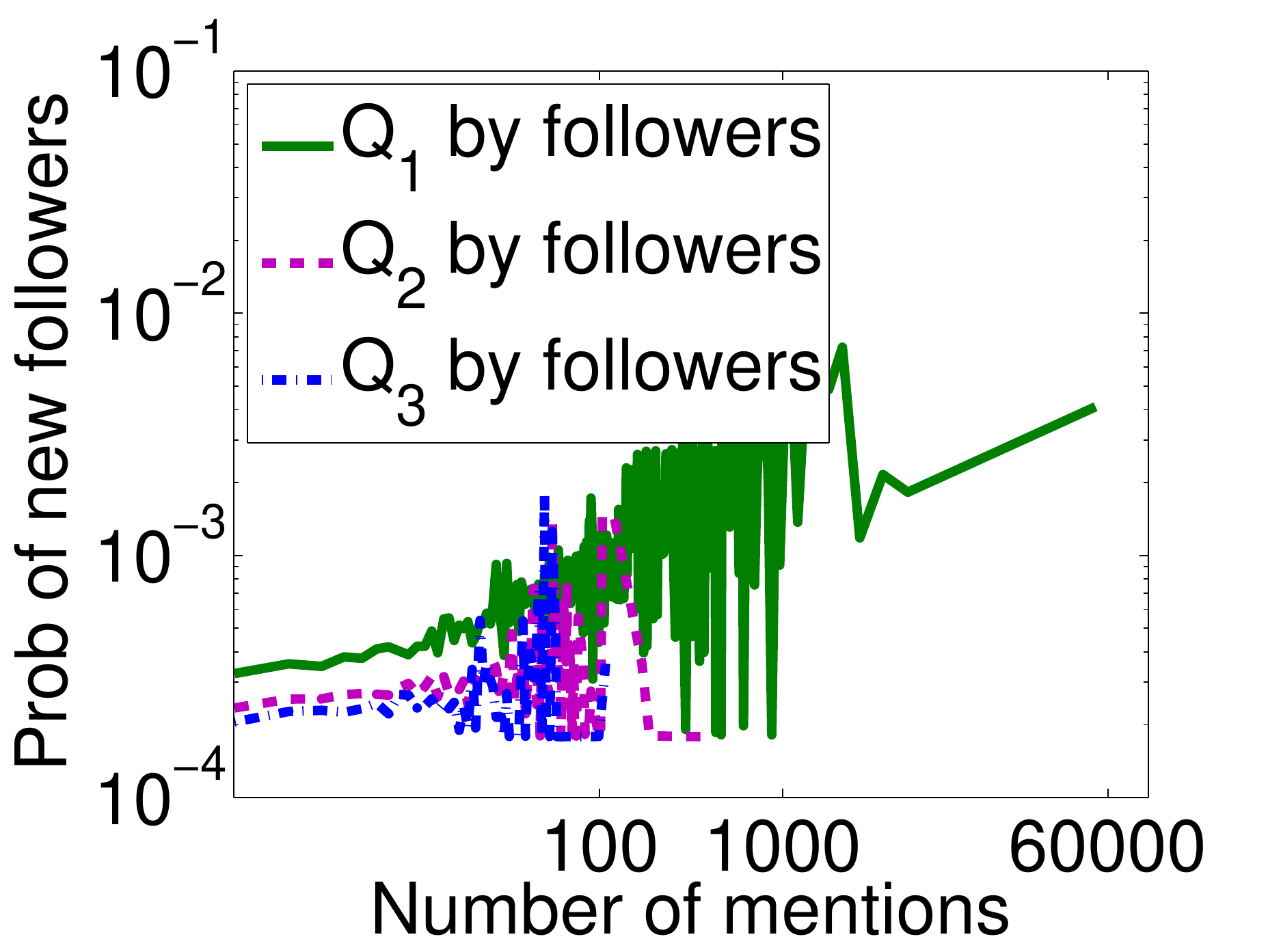}}
\caption{Probability of gaining new followers vs (a) the number of retweets and (b) mentions users receive.}\label{fig:fmr2prob}
\end{figure}

As previously observed, each retweet~\cite{Weng13kdd} and mention~\cite{Zhu14socialcom} by followers increases the attention the user receives by bringing his or her name into the feeds of other users. These others may subsequently decide to follow the mentioned or retweeted user, increasing the number of followers he or she has as a result.
Figure~\ref{subfig:R2prob2} tests this idea by plotting the mean probability of acquiring a new follower as a function of retweet count among the first three user quintiles. Generally, there is a strong positive correlation between the number of retweets and probability of getting new followers. Moreover, the correlation among users with largest number of followers is even stronger (Figure~\ref{subfig:R2prob2}) .

Figure~\ref{subfig:M2prob} shows the relationship between the number of mentions users receive and the probability of gaining new followers. Interestingly, mentions serve to attract new followers mainly for the most popular users, i.e., those with most followers.

\begin{observation}\textbf{\emph{(The Matthew effect)}}
The more attention users have, the more attention they will receive. The less attention users have, the easier it is to lose attention.
\end{observation}
\begin{figure}[!t]
{\includegraphics[width=0.98\columnwidth]{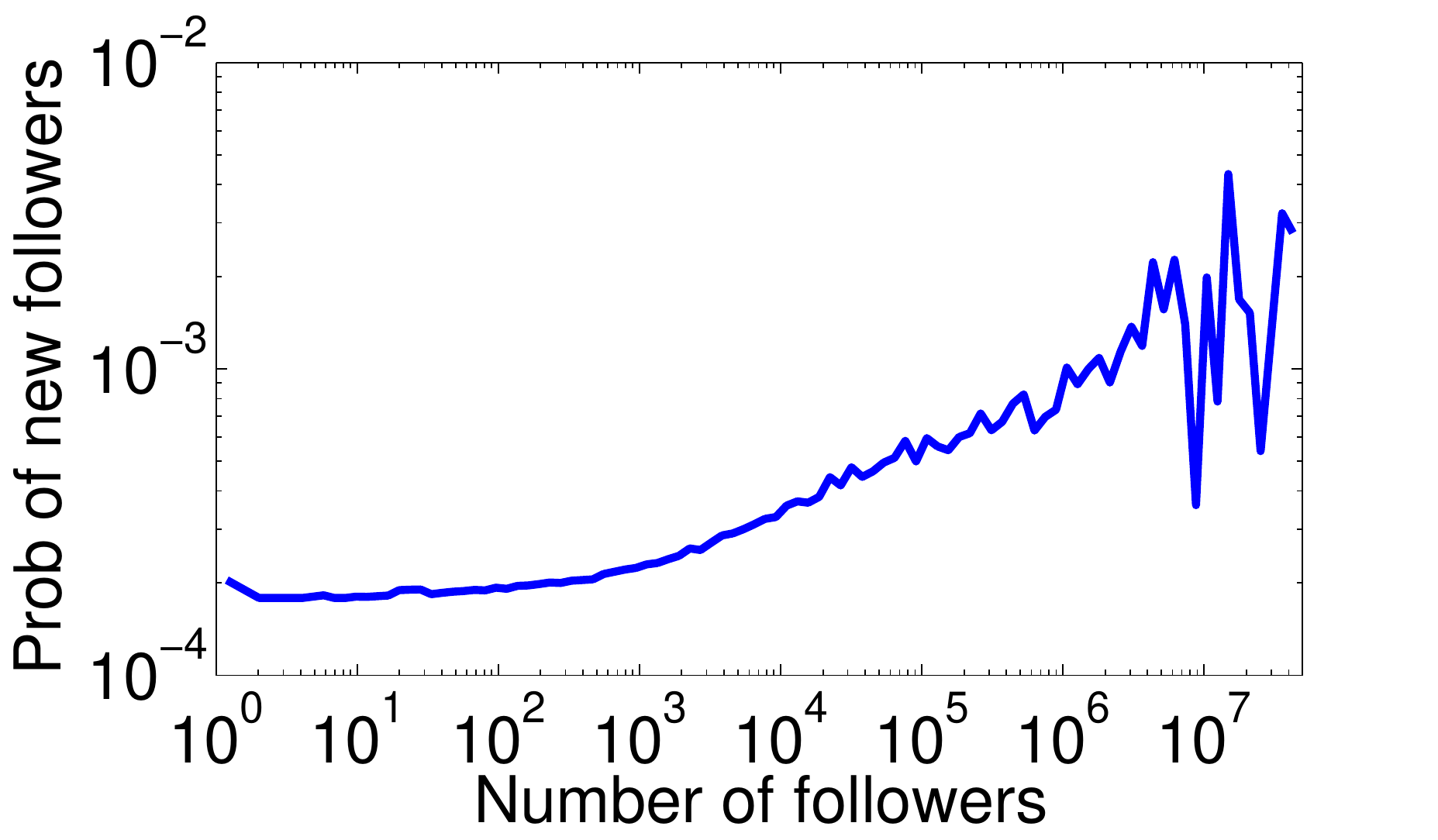}}
\caption{Probability of gaining followers vs the number of current followers users have.}
\label{subfig:f2prob}
\end{figure}

The feedback between users' popularity, measured by the number of followers they have, and the attention they receive via retweets and mentions, leads to a ``rich-get-richer'' phenomenon. Figure~\ref{subfig:f2prob} empirically validates the positive correlation between the mean probability of acquiring new follower and the number of followers users already have. Note that we log-binned the data to improve statistics; however, since there are few users with very large numbers of followers, there are large fluctuations in the extreme right portion of the plot due to poor statistics. More followers will, in turn, create more attention and even more followers. Overall, it appears that Matthew effect~\cite{Merton68} also operates in social media:  ``attention breeds attention.''


\begin{figure}[!t]
\centering
\includegraphics[width=\columnwidth]{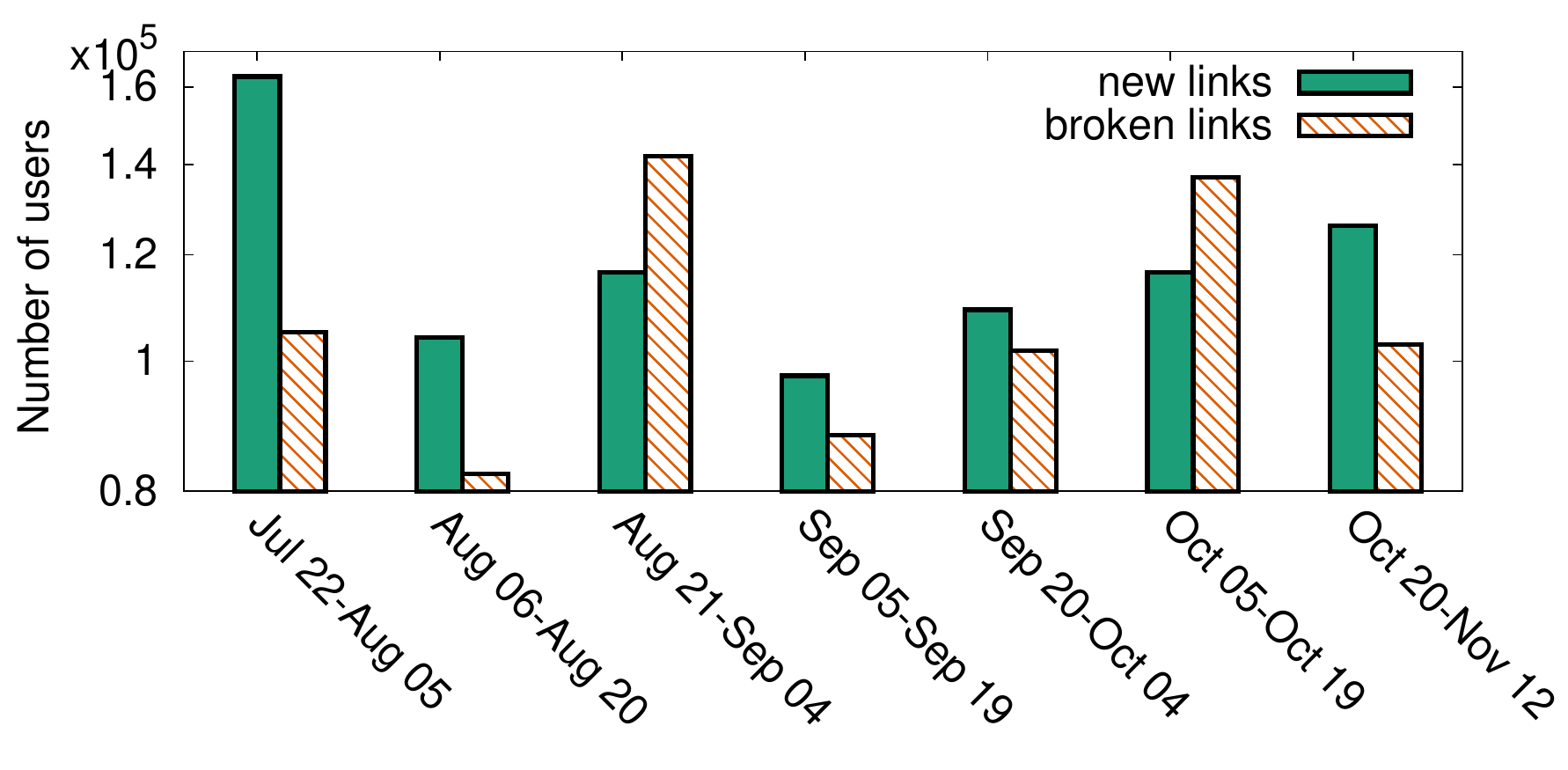}
\caption{Total number of new links and broken links made by seed users in half months.}\label{fig:brokenlinks}
\end{figure}

In addition to gaining followers, users can also lose followers in the dynamic Twitter network~\cite{MyersWWW2014}. In fact, we observed nearly as many new links created as destroyed. Figure~\ref{fig:brokenlinks} plots number of new follow links created and the number of follow links broken by seed users during each time period. Both created links and broken links contribute significantly to the dynamics of the network. The average of new friends followed per user in a two-week period is around 20.87, and average of unfollowed users per person is around 19.04. These statistics show that broken links occur as frequent as new links in social media.

\begin{figure}[!t]
  \centering
  \subfigure[]{\label{subfig:f2probminus}\includegraphics[width=0.49\columnwidth]{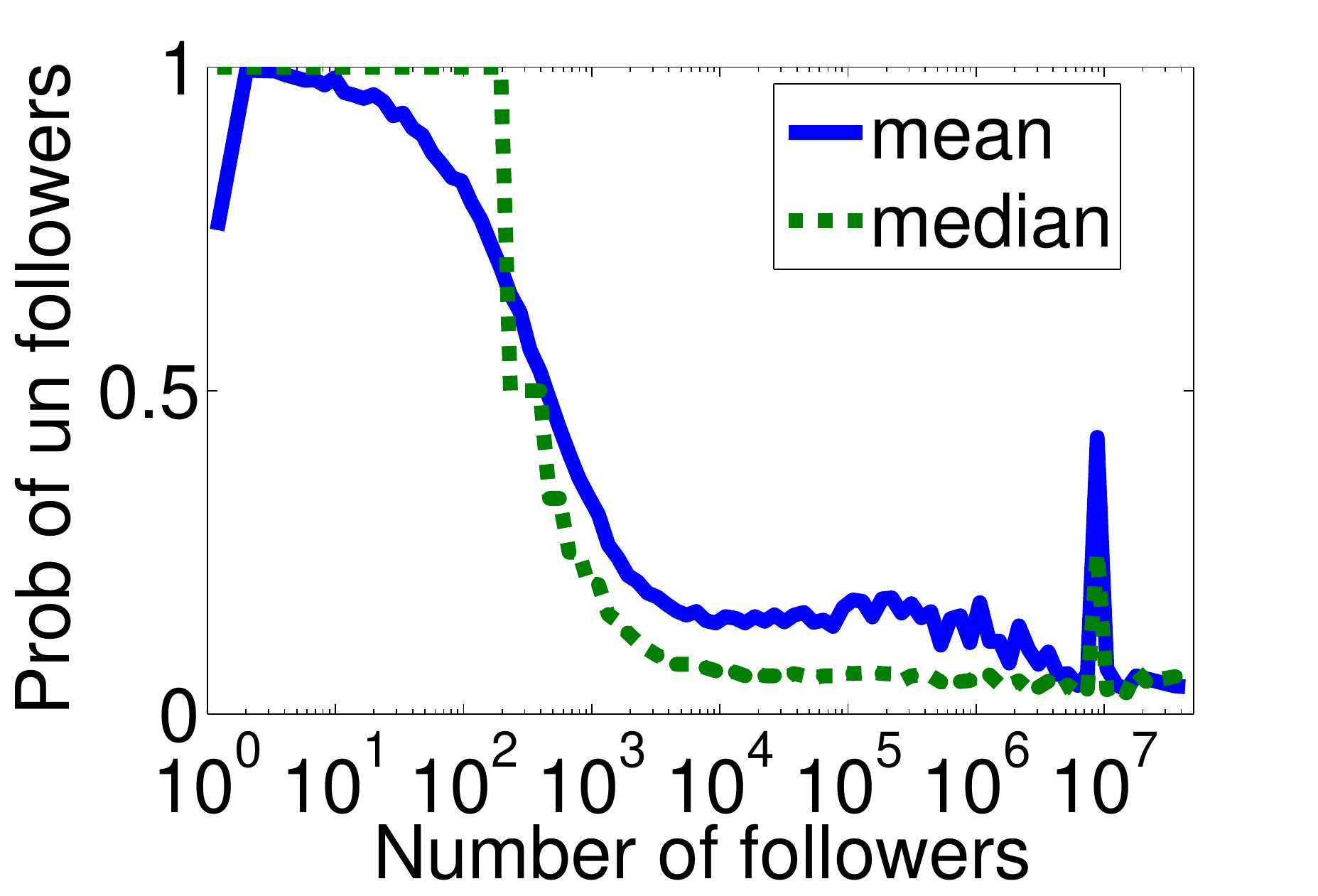}}
  \subfigure[]{\label{subfig:RT2probminus}\includegraphics[width=0.49\columnwidth]{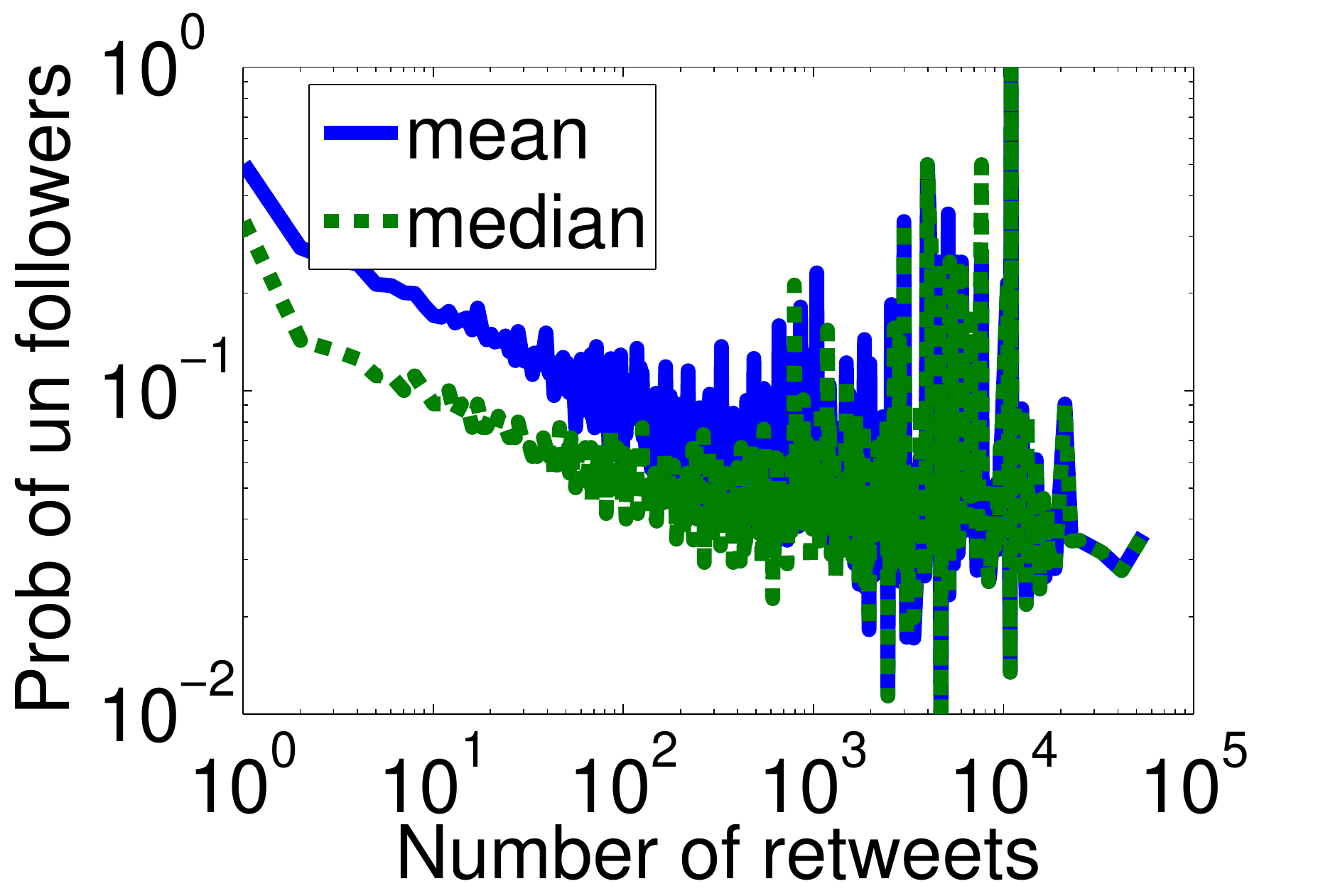}}
  \caption{ The current attention ((a) number of followers, (b) number of retweets) against the mean probability of followers they lost. Here we use logarithmic bin size.}\label{fig:fr2prob}
\end{figure}

Who loses followers and why? As shown in Figure~\ref{subfig:f2probminus}, users with many followers ($>$1,000) are far less likely to be unfollowed than users with few followers ($<$100). Unfollow probability does not vary by posting activity: for the same number of followers, users who post many status updates are not more likely to be unfollowed than users who post few status updates.

Figure~\ref{subfig:RT2probminus}  plots the probability of losing followers as a function of the number of times user's posts are retweeted. Clearly, the more times the users are retweeted, the less likely they are to lose followers, perhaps because others find their tweets valuable.




We also studied whether bursts of tweeting activity result in a loss of followers. Too many tweets appearing all at once in followers' feeds could be interpreted as spamming behavior, resulting in followers breaking the link. We adopt a Kurtosis metric to evaluate the burstiness of tweeting. Specifically, the Kurtosis metric is defined as follows:
\[\texttt{Kurt}(u)=\frac{E[(P_u-\overline{P_u})]^4}{(E[(P_u-\overline{P_u})^2])^2}\]
where $P_u$ is a random variable that denotes the number of tweets posted by $u$ at each time period, and $\overline{P_u}$ is the average number of tweets per time period.
We found, however, that tweeting burstiness does not necessarily increase the probability of either gaining new followers or losing followers. The correlation between burstiness and the probability of gaining new followers for the top 3 quin-tiles is 0.0797, -0.085 and 0.1596 respectively; while the correlation between burstiness and the probability of losing followers for the top 3 quin-tiles is 0.0765, 0.0451, and -0.0811. Therefore, for users who have low attention, burstiness of tweeting might help gain new followers. In addition, for users who have much attention, burstiness of post visibility might cause information overload and lose followers.




Collectively these observations suggest that both prongs of the Matthew effect operate to change the number of followers and increase the inequality of attention: those with most followers also acquire new followers at a faster rate, and those with the least followers also lose them at a faster rate. Since the number of followers is directly related to how much attention users receive, those rich in attention get richer, and those poor in attention get poorer.
However, users can control to some extent how much attention they receive, at least via retweeting, by increasing their posting activity without worrying that they will lose followers as a result.

\subsection{Model Description}
Attention diffusion provides a   mechanism for the evolution of the Twitter follower network. Below we introduce a dynamic model where the likelihood of forming a new follow link to a user, or breaking an existing link, depends on the amount of attention the user receives.
Thus, the main variables used by the model include the number of followers users have ($f_i$), the number of times they tweet ($p_i$), and the number of times they are mentioned ($m_i$) or their posts are retweeted ($r_i$). These variables are summarized in Table~\ref{tab:notation}.

\begin{table}
  \centering
  \caption{Notations and explanations.}\label{tab:notation}
  \begin{tabular}{|l|l|}
    \hline
    \emph{Variable} & \emph{Description}\\
    \hline \hline
    $p_i$ &number of tweets made by user $i$\\
    \hline
    $f_i$ &number of followers of $i$\\
    \hline
    $r_i$ &number of retweets of $i$'s posts by followers\\
    \hline
    $m_i$& number mentions of $i$ by followers\\
    \hline
    $P_f^+$&probability of gaining followers\\
    \hline
    $P_f^-$&probability of losing followers\\
    \hline
  \end{tabular}
\end{table}


According to the observations of the previous section, the average probability of gaining new followers $P_f^{+}$ is a function of the number of times the user is mentioned or retweeted. Figures~\ref{subfig:R2prob2} and \ref{subfig:M2prob} suggest that in our dataset this function can be approximated with
\begin{equation}
P_f^{+}\sim w_1 r^{\alpha}+w_2m^{\beta},
\end{equation}
where $w_1$ controls the contribution of retweets to the formation of new links, and $w_2$ controls the contribution of mentions to new links.

Similarly, Figure~\ref{fig:fr2prob} suggests that for our data, the average probability of unfollowing $P_f^{-}$ can be approximated with:
\begin{equation}
P_f^{-}\sim w_3 r^{-\theta}.
\end{equation}

The equations above allow us to specify how the number of followers of a user $i$ changes over time.
Let $f_i$ be a dynamic variable representing the number of followers user $i$ has at a given time. Then the rate at which user $i$ gains new followers is:
\begin{equation}\label{equ:degreeplus}
\frac{df_i^+}{dt}=(N-f_i)(w_1 r_i^{\alpha}+w_2 m_i^{\beta})
\end{equation}
where $N$ is the size of the user population, and $N-f_i$ denotes the set of available potential followers, i.e., users who do not yet follow user $i$.

We rewrite the Eq.\ref{equ:degreeplus} as
\begin{equation}\label{equ:degreeplus1}
\frac{d f_i^+}{dt}=b_if_i-b_iN
\end{equation}
where $b_i=$ $-w_1 r_i^{\alpha}-w_2 m_i^{\beta}$.
By integrating Eq.~\ref{equ:degreeplus1}, we obtain:
\begin{equation}\label{equ:degreeplus2}
f_i^{+}(t)=Ce^{b_it}-b_iNt
\end{equation}

Similarly, the rate at which user $i$ loses followers is defined as
\begin{equation}\label{equ:degreeminus}
\frac{df_i^-}{dt}=f_iw_3r_i^{-\theta}
\end{equation}
By integrating Eq.~\ref{equ:degreeminus}, we obtain:
\begin{equation}\label{equ:degreeminus2}
f_i^-(t)=C_2e^{w_3r_i^{-\theta}t}
\end{equation}
Combining Eq.~\ref{equ:degreeplus2} with Eq.~\ref{equ:degreeminus2} and using the factor that $f_i(t)=f_i(0)$ when $t=0$, we have $C_2=C$. Then the dynamic number of followers can be computed as:
\begin{equation}\label{equ:degree}
f_i(t)=f(0)+Ce^{b_it}-Ce^{w_3r_i^{-\theta}t}-b_iNt
\end{equation}

\subsection{Parameter Estimation}
The model is parameterized by constants $\alpha$, $\beta$, $\theta$, $C$, $w_1$, $w_2$, and $w_3$. We estimate these parameters using regression on data between March 20 to July 20. Since different populations of users (in different quin-tiles) might be characterized by different parameters, in the following, we separate users by quin-tile, and estimate model parameters separately for each population. We set the unit of time interval as four days. The results of estimated parameters are reported in Table~\ref{tab:parameter}.


\begin{table}
  \centering
  \caption{Model parameters estimated by regression.}\label{tab:parameter}
  \begin{tabular}{|c|c|c|c|}
    \hline
     \emph{Parameter}&\multicolumn{3}{ |c| }{\emph{User quin-tile}}\\
     \hline \hline
     &$Q_1$ & $Q_2$ & $Q_3$\\
     \hline
     $\alpha$&0.634&1.0145&0.448\\
     \hline
     $\beta$&0.865&0.0&1.141\\
      \hline
      $\theta$&0.129&-0.730&-0.020\\
      \hline
      $w_1$&0.00215&0.0&0.00006\\
      \hline
      $w_2$&0.00038&0.00030&0.0\\
      \hline
       $w_3$&0.00836&-0.00135&-300.0\\
      \hline
      $C$&8546&754&-9\\
      \hline
      RMSE&52800&406&334\\
    \hline
  \end{tabular}
\end{table}

Table~\ref{tab:parameter} indicates that different user populations attract new followers in different ways. For example, the values of parameters $\beta$ and $w_2$, which control how mentions generate new followers, are much high for users in the first quin-tile ($Q_1$) than the other quin-tiles. This shows that users with most followers, are more likely to be mentioned by others and consequently followed. The most-followed users are likely to be celebrities. When they are mentioned, and people who do not yet follow them see their name, they may recognize the name and decide to follow the celebrity in turn. For the more ordinary users in the third quin-tile $Q_3$, mentions do not lead to any new followers. For them, attention does not translate into new followers. In addition, for users in the first quin-tile, the number of followers lost decays exponentially with the number of retweets, indicating they are much less likely to lose followers if they post good content, even lots of it. In contrast, for users in the second and third quintile $Q_2$ and $Q_3$, the number of followers lost decays sub-linearly with the number of retweets. In other words, they are more likely to lose followers than first quin-tile users, even when they post content that receives similar level of attention. The attention-rich users have an advantage over the rest of the users: they accrue new followers more easily and don't lose them as much as the rest of the users.

\subsection{Solving the Model}
\begin{figure}
\centering
 \includegraphics[width=0.9\columnwidth]{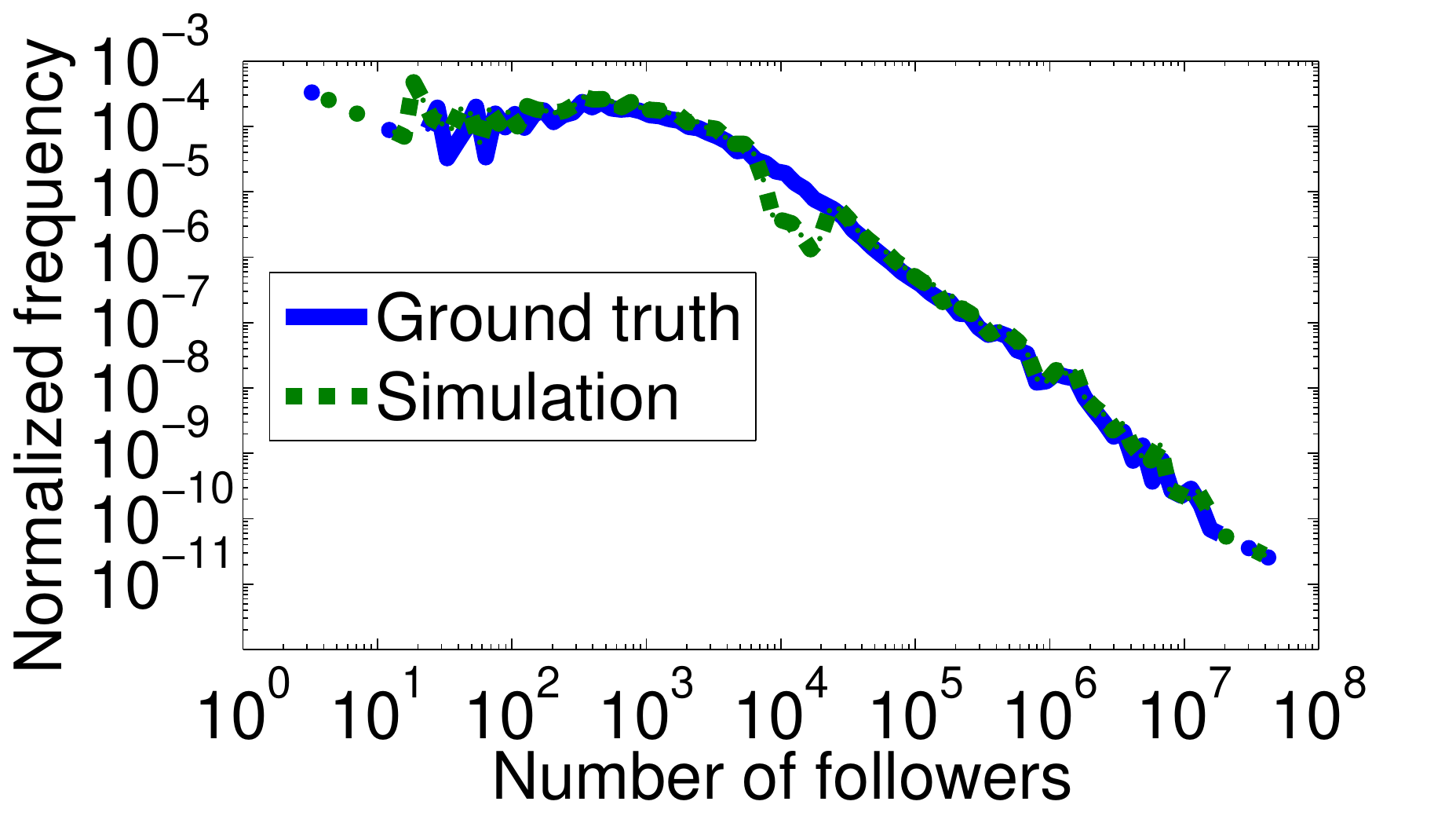}
 \caption{Distribution of the predicted and actual number of followers of seed users.}\label{fig:simF}
\end{figure}
By solving the model (Eq.~\ref{equ:degree}) with the learned parameters (Table~\ref{tab:parameter}), we can predict the number of followers each seed user in our data has at any time after training time period. Here we set the initial condition, that is, $f(0)$, as the number of followers each seed user had on July 20 2014. Figure~\ref{fig:simF}  compares the predicted distribution of followers of seed users on October 14, 2014  to the actual number of followers they have on that date (ground truth). This validates that the proposed model can correctly replicate the distribution of attention to users.

\begin{figure}
  \centering
  \includegraphics[width=\columnwidth]{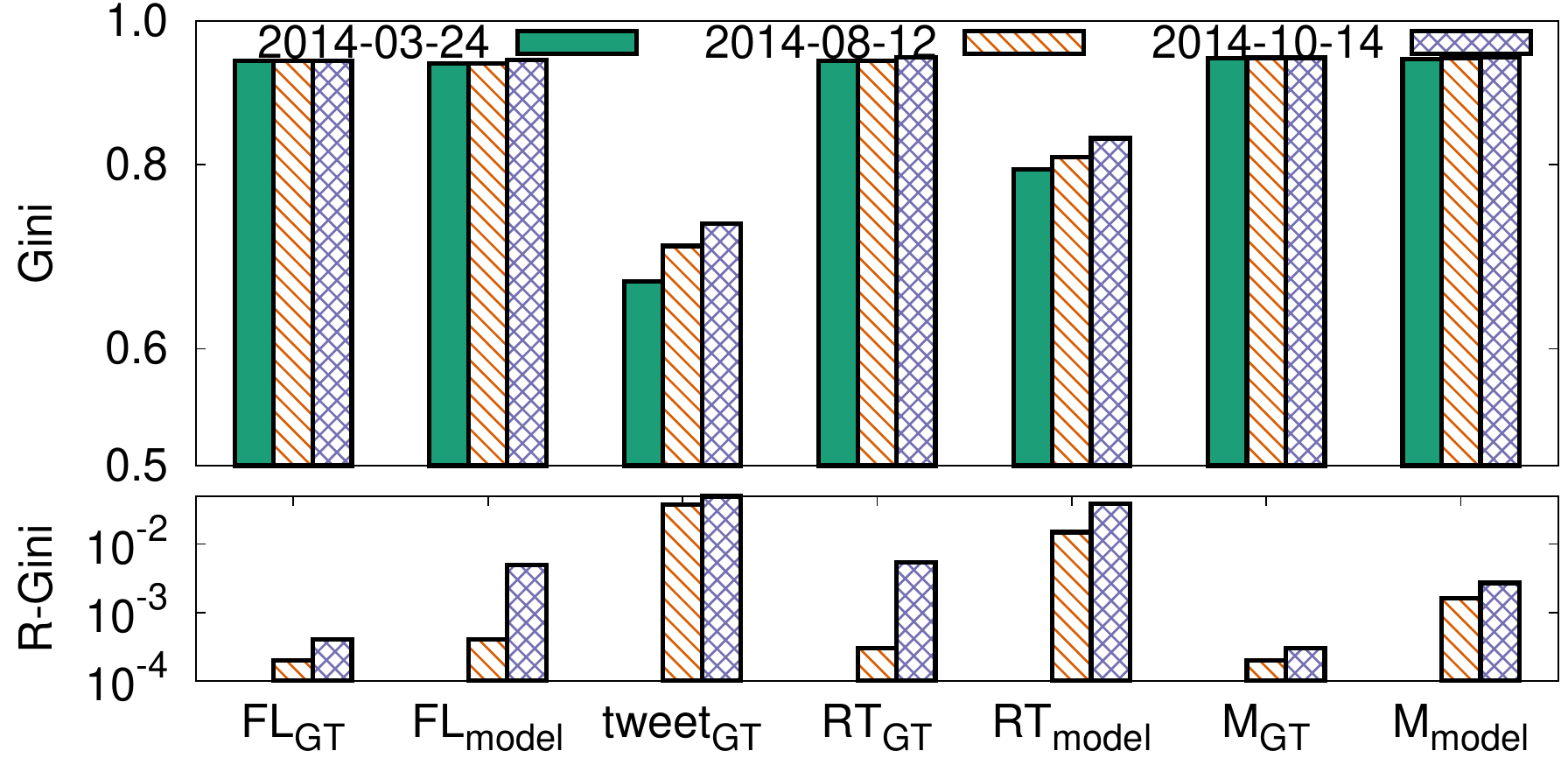}
  \caption{The Gini inequality of predicted number of followers $FL_{model}$ versus that of empirical ground truth number of followers $FL_{GT}$. In addition, the Gini of predicted number of retweets $RT_{model}$ and predicted number of mentions $M_{modeL}$ is compared with that of empirical number of retweets $RT_{GT}$ and empirical number of mentions $M_{GT}$. Since number of re-tweets is related to posting activity, we also report the inequality of posting activity during one week interval $tweet_{GT}$. }\label{fig:simGini}
\end{figure}

Next, we examine how well the proposed model predicts dynamics of follower inequality. Figure~\ref{fig:simGini} compares the inequality of the predicted and actual number of followers at three different points in time. The attention diffusion model results in high follower inequality, with values of the Gini coefficient around 0.94. In addition, in the attention diffusion model, inequality grows over time as attention becomes more concentrated on a small number of users. The trends using the proposed attention model are close to the empirical ground truth. 

We can also use the attention diffusion model to predict inequality of attention via retweets over time. We model the number of retweets users receive is a function of the number of followers they have and their tweeting activity, and then predict the number of retweets using regression. Specifically, the number of retweets a user received, can be approximately modeled as:

\begin{equation}\label{equ:retweetmodel}
r_i=a f_i^b + c p_i^d
\end{equation}

We can then predict the number of retweets users get at any time by using their predicted number of followers and their observed tweeting activity at that time.
As shown in Figure~\ref{fig:simGini}, the Gini coefficient of the predicted number of retweets rises over time; however, it is much less than the ground truth value. The possible reason is that in Eq.~\ref{equ:retweetmodel}, the number of retweets is directly related to tweeting activity . Since tweeting activity has much lower inequality, the inequality of number of predicted retweets is also lower than the ground truth value.

Similar to above process, we model the number of mentions users received as a function of number of followers:
\begin{equation}\label{equ:mentionmodel}
m_i=a e^{b f_i}
\end{equation}

With Eq.~\ref{equ:mentionmodel}, we predict the number of mentions at three different time points and report its Gini coefficient in Figure~\ref{fig:simGini}. The Gini coefficient of the predicted value, is very close to the Gini coefficient of number of followers and the ground truth. However, the marginal differences between two different time points in the ground truth, are relatively smaller than those in the prediction. This is due to in Eq.~\ref{equ:mentionmodel}, number of followers is the only dependant factor of number of mentions.

In conclusion, the proposed model describes attention diffusion among users in a network and can successfully predict the number of future followers each user will have and the resulting attention inequality. However, it is unclear how to use it to model the attention to messages being tweeted. Besides, results in Figure~\ref{fig:simGini} show that the attention user received in terms of the number of retweets and the number of mentions, may not simply be modeled by Eq.~\ref{equ:retweetmodel} and Eq.~\ref{equ:mentionmodel}. One of the future direction is to model attention diffusion in terms of number of retweets and number of mentions with both network and content information.

\section{Related Works}\label{sec-related}
Our work touches on topics from a variety of fields. We briefly review some of the relevant literature.

\subsection{Inequality}
The study of economic inequality has a long history beginning with the quantitative study of personal wealth distribution by Vilfredo Pareto~\cite{pareto1896}. In this work, the empirical data analysis showed that the tail of wealth distribution follows a power-law. More recent analysis showed that wealth inequality (in the U.S.) is not only large~\cite{Ariely11inequality} but increasing with time~\cite{Piketty14science}. A number of models~\cite{gcalda:Champernowne53,Scafetta2004,Silva05temporalevolution} have been suggested to understand the features of empirical wealth distributions and relate them to appropriate mechanisms. For example, Banerjee and Yakovenko~\cite{Banerjee10} adopted the Fokker-Planck equation model for the income distribution, which leads to an approximated exponential for small and mid-range incomes, and as a power-law for the highest incomes.

Inequality exists everywhere online~\cite{Wilkinson08}. On the World Wide Web, 80\% of the browsing traffic goes to 5\% of the hostnames~\cite{Webineqality}. The browsing distribution is also highly skewed among URLs, sites and domains. Fortunato et al.~\cite{Fortunato06prl} suggests that search engines bias user traffic by their page ranking strategies. They observed a high correlation between click probability and hit rank. Therefore, the ranking process has a potential to skew browsing attention to those top ranked URLs.

Recently, with the advance of social networks, the studies of socio-economic inequalities have sparked a renewed interests from researchers. Specifically, much research has focused on a better understanding of the nature and origin of inequalities from user behavior and network aspects. Among them, Fuchs and Thurner~\cite{Wealthinequality2014} analyzed data from the virtual economy of the massive multi-player online game, to find the explanations for origins of wealth inequality in terms of behavior and network. Their results suggested that wealthy players have high in- and out-degree in the trade network; while in contrast, players that are not socially well-connected are significantly poorer on average. In addition to wealth inequality, Chatterjee~\cite{Socio-economic2014} have also studied the city size inequality, firm inequality, voting inequality, opinion inequality and their origins from a statistical perspective. In this work, we aim to study the attention inequalities in social media and to understand some possible reasons that lead to the inequalities. In addition, we are also interested in understanding how inequalities changes over time.

\subsection{Network Growth}
Nodel et al.~\cite{Noelunfriend2011} analyzed the ``unfriending" behavior using longitudinal social network data. Their results indicate that if there is a non-negligible friendship attrition over time, homophily in friendship retention induces significant upward bias to peer influence. Kwak~\cite{KwakCHI2011} studied the dynamics of unfollowing in Twitter and showed that the reciprocity of the relationships, the duration of a relationship, the followee's informativeness, and the overlap of the relationships, affect the decision to unfollow. Myers and Leskovec~\cite{MyersWWW2014} studied the bursty dynamics of the Twitter Information Network and found that the dynamics of network structure can be characterized by steady rates of change but with interruption by sudden bursts. They also developed a model that quantifies the dynamics of the network and predicts the bursty of information diffusion events.

An array of models for describing the growth of social networks have been described in literature. \emph{Random Walk Model}~\cite{PAmodel} was introduced to model the randomized process of creating new links in a social network.  The \emph{BA Model}~\cite{Barabasi99emergenceScaling} uses the preferential attachment mechanism~\cite{PAmodel} to generate the rich-get-richer phenomenon. The \emph{Nearest Neighbor Model}~\cite{Newman01clusteringand} was proposed based on the observation that the probability of two unacquainted users to become friends increases with the (weighted) number of common friends. Durr et al.~\cite{DurrASONAM2012} proposed a new model based on homophily for both P2P and centralized online social network. Their model is able to explain the distribution and frequency of interactions in online social networks. In our attention model, we aim to characterize the attention diffusion and network dynamic mechanisms. In addition, we attempt for explanations for inequalities and inequality dynamics from our attention model.

\section{Conclusion}
We analyzed how people allocate their attention to other users on a popular social media site Twitter by following them, sharing (retweeting) the messages they post, and mentioning them in their own posts. We concluded that attention is non-uniformly distributed, with a small number of users dominating the attention received. This results in a high attention inequality, much higher than even the widely discussed economic inequalities. Moreover, we showed that attention inequality is increasing on Twitter.

To explain the rise of attention inequality, we studied how the structure of the follower network changes due to attention diffusion. We showed empirically that the attention users receive by being mentioned or having their messages retweeted leads to them gain new followers. The more attention users have, the less likely they are to lose followers, resulting in a ``rich get richer, poor get poorer'' dynamic. We constructed a phenomenological model of network dynamics based on these observations, and used the model to predict the number of followers a user will have in the future, and estimate future levels of attention inequality. In this model, attention inequality grows in time.



Our work leaves many questions unanswered. What are the benefits of inequality? Is it due to easier to deal with information overload if users only have to pay attention to popular items? What are the disadvantages of inequality? Is it associated with less diversity in content and viewpoints online? What motivates people to be engaged and contribute to social media when they do not receive any attention? We plan to address some of these questions in future work.

\bibliographystyle{aaai}
\bibliography{Inequality}

\begin{thebibliography}{}

\bibitem[\protect\citeauthoryear{Adler}{1985}]{Adler}
Adler, M.
\newblock 1985.
\newblock Stardom and talent.
\newblock {\em The American Economic Review} 75(1):208--212.

\bibitem[\protect\citeauthoryear{Allison, Long, and Kraze}{1982}]{Allison82}
Allison, P.~D.; Long, J.~S.; and Kraze, T.~K.
\newblock 1982.
\newblock {Cumulative advantage and inequality in science}.
\newblock {\em Ame. Sociological Review} 47(5):615--625.

\bibitem[\protect\citeauthoryear{Allison}{1980}]{Allison80}
Allison, P.~D.
\newblock 1980.
\newblock {Inequality and Scientific Productivity}.
\newblock {\em Social Studies of Science} 10(2):163--179.

\bibitem[\protect\citeauthoryear{Banerjee and Yakovenko}{2010}]{Banerjee10}
Banerjee, A., and Yakovenko, V.~M.
\newblock 2010.
\newblock Universal patterns of inequality.
\newblock {\em New Journal of Physics} 12(7):075032.

\bibitem[\protect\citeauthoryear{Barabasi and
  Albert}{1999}]{Barabasi99emergenceScaling}
Barabasi, A.-L., and Albert, R.
\newblock 1999.
\newblock Emergence of scaling in random networks.
\newblock {\em Science} 286(5439):509--512.

\bibitem[\protect\citeauthoryear{Champernowne}{1953}]{gcalda:Champernowne53}
Champernowne, D.
\newblock 1953.
\newblock A model for income distribution.
\newblock {\em Economic Journal} 63:318--351.

\bibitem[\protect\citeauthoryear{Chatterjee}{2014}]{Socio-economic2014}
Chatterjee, A.
\newblock 2014.
\newblock Socio-economic inequalities: a statistical physics perspective.
\newblock Technical report, arXiv.org.

\bibitem[\protect\citeauthoryear{Durr, Protschky, and
  Linnhoff-Popien}{2012}]{DurrASONAM2012}
Durr, M.; Protschky, V.; and Linnhoff-Popien, C.
\newblock 2012.
\newblock Modeling social network interaction graphs.
\newblock In {\em IEEE/ACM International Conference on Advances in Social
  Networks Analysis and Mining (ASONAM)},  660--667.

\bibitem[\protect\citeauthoryear{Fortunato, Flammini, and
  Menczer}{2006}]{Fortunato06prl}
Fortunato, S.; Flammini, A.; and Menczer, F.
\newblock 2006.
\newblock Scale-free network growth by ranking.
\newblock {\em Phys. Rev. Lett.} 96:218701.

\bibitem[\protect\citeauthoryear{Fuchs and
  Thurner}{2014}]{Wealthinequality2014}
Fuchs, B., and Thurner, S.
\newblock 2014.
\newblock Behavioral and network origins of wealth inequality: Insights from a
  virtual world.
\newblock {\em PLoS ONE} 9(8):e103503.

\bibitem[\protect\citeauthoryear{Gini}{1997}]{gini_concentration_1997}
Gini, C.
\newblock 1997.
\newblock Concentration and dependency ratios.
\newblock {\em Rivista di Politica Economica} 87:769--792.

\bibitem[\protect\citeauthoryear{Hodas, Kooti, and Lerman}{2013}]{Hodas13icwsm}
Hodas, N.~O.; Kooti, F.; and Lerman, K.
\newblock 2013.
\newblock Friendship paradox redux: Your friends are more interesting than you.
\newblock In {\em Proceedings of the 7Th International Aaai Conference On
  Weblogs And Social Media (ICWSM)}.

\bibitem[\protect\citeauthoryear{Kawachi and Kennedy}{1999}]{Kawachi99health}
Kawachi, I., and Kennedy, B.~P.
\newblock 1999.
\newblock Income inequality and health: pathways and mechanisms.
\newblock {\em Health services research} 34(1 Pt 2):215--227.

\bibitem[\protect\citeauthoryear{Kelly}{2000}]{Kelly00crime}
Kelly, M.
\newblock 2000.
\newblock Inequality and crime.
\newblock {\em Review of Economics and Statistics} 82(4):530--539.

\bibitem[\protect\citeauthoryear{Klamer and Van~Dalen}{2002}]{Klamer02}
Klamer, A., and Van~Dalen, H.~P.
\newblock 2002.
\newblock Attention and the art of scientific publishing.
\newblock {\em Journal of Economic Methodology} 9(3):289--315.

\bibitem[\protect\citeauthoryear{Kwak, Chun, and Moon}{2011}]{KwakCHI2011}
Kwak, H.; Chun, H.; and Moon, S.
\newblock 2011.
\newblock Fragile online relationship: A first look at unfollow dynamics in
  twitter.
\newblock In {\em Proceedings of the SIGCHI Conference on Human Factors in
  Computing Systems},  1091--1100.
\newblock ACM.

\bibitem[\protect\citeauthoryear{Lariviere and Gingras}{2010}]{Lariviere09}
Lariviere, V., and Gingras, Y.
\newblock 2010.
\newblock {The impact factor's Matthew effect: a natural experiment in
  bibliometrics}.
\newblock {\em Journal of the American Society for Information Science and
  Technology} 61(2):424--427.

\bibitem[\protect\citeauthoryear{McCurley}{2007}]{Webineqality}
McCurley, K.
\newblock 2007.
\newblock Income inequality in the attention economy.
\newblock {\em http://mccurley.org/papers/effective}.

\bibitem[\protect\citeauthoryear{Merton}{1968}]{Merton68}
Merton, R.~K.
\newblock 1968.
\newblock {The Matthew Effect in Science}.
\newblock {\em Science} 159(3810):56--63.

\bibitem[\protect\citeauthoryear{Myers and Leskovec}{2014}]{MyersWWW2014}
Myers, S.~A., and Leskovec, J.
\newblock 2014.
\newblock The bursty dynamics of the twitter information network.
\newblock In {\em Proceedings of the 23rd International Conference on World
  Wide Web},  913--924.
\newblock ACM.

\bibitem[\protect\citeauthoryear{Newman}{2001}]{Newman01clusteringand}
Newman, M. E.~J.
\newblock 2001.
\newblock Clustering and preferential attachment in growing networks.
\newblock {\em Phys. Rev. E}.

\bibitem[\protect\citeauthoryear{Noel and Nyhan}{2011}]{Noelunfriend2011}
Noel, H., and Nyhan, B.
\newblock 2011.
\newblock The ??? unfriending??? problem: The consequences of homophily in
  friendship retention for causal estimates of social influence.
\newblock {\em Social Networks} 33(3):211--218.

\bibitem[\protect\citeauthoryear{Norton and Ariely}{2011}]{Ariely11inequality}
Norton, M.~I., and Ariely, D.
\newblock 2011.
\newblock Building a better {America?One} wealth quintile at a time.
\newblock {\em Perspectives on Psychological Science} 6(1):9--12.

\bibitem[\protect\citeauthoryear{Pareto}{1896}]{pareto1896}
Pareto, V.
\newblock 1896.
\newblock {\em Cours d'Economie Politique}.

\bibitem[\protect\citeauthoryear{Piketty and Saez}{2014}]{Piketty14science}
Piketty, T., and Saez, E.
\newblock 2014.
\newblock {Inequality in the long run}.
\newblock {\em Science} 344(6186):838--843.

\bibitem[\protect\citeauthoryear{Salganik, Dodds, and Watts}{2006}]{Salganik06}
Salganik, M.; Dodds, P.; and Watts, D.
\newblock 2006.
\newblock {Experimental Study of Inequality and Unpredictability in an
  Artificial Cultural Market}.
\newblock {\em Science} 311(5762):854--856.

\bibitem[\protect\citeauthoryear{Scafetta, Picozzi, and
  West}{2004}]{Scafetta2004}
Scafetta, N.; Picozzi, S.; and West, B.~J.
\newblock 2004.
\newblock An out-of-equilibrium model of the distributions of wealth.
\newblock {\em Quantitative Finance} 4(3):353--364.

\bibitem[\protect\citeauthoryear{Silva and
  Yakovenko}{2005}]{Silva05temporalevolution}
Silva, A.~C., and Yakovenko, V.~M.
\newblock 2005.
\newblock Temporal evolution of the ???thermal??? and ???superthermal??? income
  classes in the usa during 1983--2001.
\newblock {\em Europhysics Letters}.

\bibitem[\protect\citeauthoryear{V\'azquez}{2003}]{PAmodel}
V\'azquez, A.
\newblock 2003.
\newblock Growing network with local rules: Preferential attachment, clustering
  hierarchy, and degree correlations.
\newblock {\em Phys. Rev. E} 67:056104.

\bibitem[\protect\citeauthoryear{Weng \bgroup et al\mbox.\egroup
  }{2013}]{Weng13kdd}
Weng, L.; Ratkiewicz, J.; Perra, N.; Gon\c{c}alves, B.; Castillo, C.; Bonchi,
  F.; Schifanella, R.; Menczer, F.; and Flammini, A.
\newblock 2013.
\newblock The role of information diffusion in the evolution of social
  networks.
\newblock In {\em Proceedings of the 19th ACM SIGKDD International Conference
  on Knowledge Discovery and Data Mining},  356--364.

\bibitem[\protect\citeauthoryear{Wilkinson}{2008}]{Wilkinson08}
Wilkinson, D.~M.
\newblock 2008.
\newblock Strong regularities in online peer production.
\newblock In {\em EC '08: Proceedings of the 9th ACM conference on Electronic
  commerce},  302--309.
\newblock New York, NY, USA: ACM.

\bibitem[\protect\citeauthoryear{Zhu and Lerman}{2014}]{Zhu14socialcom}
Zhu, L., and Lerman, K.
\newblock 2014.
\newblock A visibility-based model for link prediction in social media.
\newblock In {\em Proceedings of the ASE/IEEE Conference on Social Computing}.

\end{thebibliography}
\end{document}